\documentclass[pra,letterpaper,aps,10pt,superscriptaddress,twocolumn,floatfix,showpacs]{revtex4-1}
\usepackage{graphicx}
\usepackage{amsmath}
\usepackage{amsfonts}
\usepackage{float}
\usepackage{amssymb}
\usepackage{bibunits}
\usepackage{epsfig}
\usepackage{epstopdf}
\DeclareGraphicsExtensions{.pdf,.eps,.png,.jpg,.mps}
\usepackage[pdftex]{color}
\usepackage{amsmath,graphicx,amssymb,braket,xcolor,subfigure,upgreek}
\usepackage[colorlinks, linkcolor=blue, citecolor=blue, urlcolor=blue, breaklinks=true]{hyperref}
\usepackage{microtype}
\usepackage{mathtools}
\usepackage{bbm}
\usepackage{color}

\newcommand{\R}{\text{R}}
\newcommand{\Le}{\text{L}}

\newcommand{\omegamax}{\omega_{\text{max}}}
\newcommand{\Gammam}{\Gamma_{\text{m}}}

\begin{document}

\title{
Molecule-photon interactions in phononic environments}
\author{M.~Reitz}
\affiliation{Max Planck Institute for the Science of Light, Staudtstra{\ss}e 2,
D-91058 Erlangen, Germany}
\affiliation{Department of Physics, University of Erlangen-Nuremberg, Staudtstra{\ss}e 7,
D-91058 Erlangen, Germany}

\author{C.~Sommer}
\affiliation{Max Planck Institute for the Science of Light, Staudtstra{\ss}e 2,
D-91058 Erlangen, Germany}

\author{B.~Gurlek}
\affiliation{Max Planck Institute for the Science of Light, Staudtstra{\ss}e 2,
D-91058 Erlangen, Germany}
\affiliation{Department of Physics, University of Erlangen-Nuremberg, Staudtstra{\ss}e 7,
D-91058 Erlangen, Germany}

\author{V.~Sandoghdar}
\affiliation{Max Planck Institute for the Science of Light, Staudtstra{\ss}e 2,
D-91058 Erlangen, Germany}
\affiliation{Department of Physics, University of Erlangen-Nuremberg, Staudtstra{\ss}e 7,
D-91058 Erlangen, Germany}
\date{\today}
\date{\today}

\author{D.~Martin-Cano}
\affiliation{Max Planck Institute for the Science of Light, Staudtstra{\ss}e 2,
D-91058 Erlangen, Germany}
\date{\today}

\author{C.~Genes}
\affiliation{Max Planck Institute for the Science of Light, Staudtstra{\ss}e 2,
D-91058 Erlangen, Germany}
\affiliation{Department of Physics, University of Erlangen-Nuremberg, Staudtstra{\ss}e 7,
D-91058 Erlangen, Germany}
\date{\today}
\date{\today}

\begin{abstract}
Molecules constitute compact hybrid quantum optical systems that can interface photons, electronic degrees of freedom, localized mechanical vibrations and phonons. In particular, the strong vibronic interaction between electrons and nuclear motion in a molecule resembles the optomechanical radiation pressure Hamiltonian. While molecular vibrations are often in the ground state even at elevated temperatures, one still needs to get a handle on decoherence channels associated with phonons before an efficient quantum optical network based on opto-vibrational interactions in solid-state molecular systems could be realized. As a step towards a better understanding of decoherence in phononic environments, we take here an open quantum system approach to the non-equilibrium dynamics of guest molecules embedded in a crystal, identifying regimes of Markovian versus non-Markovian vibrational relaxation. A stochastic treatment based on quantum Langevin equations predicts collective vibron-vibron dynamics that resembles processes of sub- and superradiance for radiative transitions. This in turn leads to the possibility of decoupling intramolecular vibrations from the phononic bath, allowing for enhanced coherence times of collective vibrations. For molecular polaritonics in strongly confined geometries, we also show that the imprint of opto-vibrational couplings onto the emerging output field results in effective polariton cross-talk rates for finite bath occupancies.
\end{abstract}


\maketitle

\section{Introduction}

Molecules are natural quantum mechanical platforms where several atoms are interlinked via electronic bonds. The inherent coupling between the electronic transitions at optical frequencies and the mechanical nuclear motions (vibrons) at terahertz frequencies renders molecular systems ideal for the realization of quantum optomechanical effects. This is however different from the radiation pressure coupling mechanism in macroscopic systems, as optomechanical interactions in molecules intrinsically occur in a hybrid fashion involving a two-step process of photon-electron and vibron-electron (vibronic) interactions \cite{moerner2002adozen, roelli2016molecular, benz2016single, neuman2019quantum}. The vibronic coupling resembles the radiation pressure Hamiltonian (via a boson-spin replacement) which can be in the strong coupling regime since the strength of the coherent coupling can be comparable to the vibrational frequency. At cryogenic temperatures (e.g., at $T\sim 4\,\mathrm{K}$), molecular vibrations are in their quantum ground state thus circumventing usual complications arising from additional optical cooling requirements \cite{mahler1996molecular}. Moreover, naturally occurring or engineered differences in the curvatures of the ground and excited state potential surfaces of the molecular electronic orbitals can lead to the direct generation of non-classical squeezed vibrational wavepackets \cite{averbukh1993optimal}. These aspects suggest that molecular systems offer natural platforms, where one can exploit the inherent opto-vibrational coupling as a quantum resource.\\
%\indent When molecules couple to their environment, e.g. in the solid state, the mechanical modes of localized intramolecular vibrations (vibrons) are augmented by collective delocalized vibrational excitations of the host material (phonons), which allow for electron-phonon (polaron) couplings. These aspects suggest that molecular systems offer natural platforms, where one can exploit the inherent opto-vibrational coupling as a quantum resource.\\
\indent When molecules couple to their condensed-matter environment, e.g.~in the solid state, the mechanical modes of localized intramolecular vibrations (vibrons) are augmented by collective delocalized vibrational excitations of the host material (phonons), which allow for electron-phonon (polaron) couplings. In practice, coupling to a large number of phonon modes makes the study of molecular vibrations in the solid state notoriously challenging. Some of the challenges can be tamed under cryogenic conditions where experiments manage to reduce phonon coupling on the so-called zero-phonon line (ZPL) of the transition between $\ket{g, n_\nu=0}$ and $\ket{e, n_\nu=0}$ sufficiently to reach its natural linewidth limit. This can be verified in ensemble measurements, e.g.~via hole burning, or in single-molecule spectroscopy \cite{rigler2001single}. A good example of an experimental platform is provided by dibenzoterrylene (DBT) molecules embedded in anthracene crystals [see Fig.~\ref{fig1}(a)], exhibiting a lifetime-limited linewidth and near-unity radiative yields at cryogenic temperatures \cite{nicolet2007single, kozankiewicz2014single, lombardi2018photostable, polisseni2016stable, wang2019turning}. However, even if vibrational spectroscopy at the single-molecule level is readily accessible in the laboratory \cite{makarewicz2012vibronic, deperasinska2011single}, a quantitative understanding of the couplings between the molecular vibrational modes and their internal and external degrees of freedom is still largely missing. In particular, a detailed study of decoherence sources is necessary.\\
\indent An open quantum system approach, such as employed in our treatment, can shed light onto a few aspects of coherent and incoherent vibrational dynamics and onto the light-matter interactions in the presence of vibrons, phonons and cavity-localized photon modes. Our formalism makes use of quantum Langevin equations which allows us to follow the evolution of system operators such as the electronic coherence and vibrational quadratures and to derive analytical results for the time dynamics of both expectation values and two-time correlations (needed for the computation of emission and absorption spectra). We find that closely spaced molecules can experience collective vibrational relaxation, an effect similar to the sub- and superradiance of quantum emitters in the electromagnetic vacuum. This can be exploited to decouple collective two-molecule vibrational states from the decohering phononic environment leading to the possibility of coherently mapping motion onto light and vice versa. In addition, at the level of the pure light-matter interface, coupling to confined optical cavity modes can increase the oscillator strength of the molecule by effectively reducing vibronic couplings \cite{wang2019turning}.

\indent Our formalism also allows us to treat problems relevant to experiments in cavity quantum electrodynamics with molecules, where standard concepts such as strong coupling or the Purcell effect can suffer important modifications once couplings between electronic transitions and vibrations are taken into account. To this end, we make use of analytical tools based on quantum Langevin equations \cite{reitz2019langevin} to account for an arbitrary number of vibron and phonon modes. Earlier theoretical works have either traced out the typically fast vibrational degrees of freedom \cite{wang2017coherent, haakhsqueezed2015}, used limited numerical simulations, or focused mostly on aspects such as vibrational relaxation in solids \cite{rebane1970impurity,bondybev1984relaxation, knox2002low, hill1988vibrational}, electron-phonon and electron-vibron couplings \cite{sild1988zero, hochstrasser1972phonon, vogel1986theory}, temperature dependence of the zero-phonon linewidth \cite{mccumber1963linewidth, silsbee1962thermal} and anharmonic effects \cite{kenkre1994theory, waxer1997molecular}.

However, it should also be borne in mind that the relevance of our treatment is not restricted to the physical system considered here as very similar effects also occur in related solid-state emitters such as quantum dots or vacancy centers in diamond. The coupling of such systems to photonic nanostructures has been studied quite extensively over the last years \cite{ilessmith2017limits, norambuena2016microscopic, kaer2010nonmarkovian, mccutcheon2010quantum, nazir2016modeling, lodahl2015interfacing, brash2019light,englund2010deterministic}. There is, furthermore, a general current interest in impurities interacting with a quantum many-body environment, such as molecular rotors immersed in liquid solvents \cite{lemeshko2015rotation, lemeshko2017quasiparticle}, Rydberg impurities in quantum gases \cite{schmidt2016mesoscopic} or magnetic polarons in the Fermi-Hubbard model \cite{koepsell2019imaging}. Our treatment can then be understood as a general model for the coupled dynamics of spin systems to  many, possibly interconnected, bosonic degrees of freedom as illustrated in Fig.~\ref{fig1}(c).

%These features makes solid-state embedded molecules promising elements for integrated photonic circuitry, e.g.~as single photon sources \cite{lombardi2018photostable, polisseni2016stable, brunel1999triggered, trebbia2009efficient} or in providing optical nonlinearities \cite{maser2016few,wang2019turning}. The aim is to establish a building block towards the construction of a quantum network where each molecule can provide a coherent link between light and matter. From a more fundamental point of view, optical spectroscopy at the single-molecule level in the condensed phase allows to study basic optical properties of molecules which would otherwise not be accessible or get lost in ensemble measurements due to inhomogeneities \cite{moerneroptical1989,moerner1999illuminating}. While it is well known that the immersion into a solid-state environment changes the optical properties of molecules, an analytical theoretical understanding of the interplay between the underlying mechanisms in these laser-driven dissipative systems is still lacking, especially regarding the coupling to confined optical modes.

\section{Model}
\label{model}

\subsection{General considerations}
\begin{figure}[t]
\includegraphics[width=0.96\columnwidth]{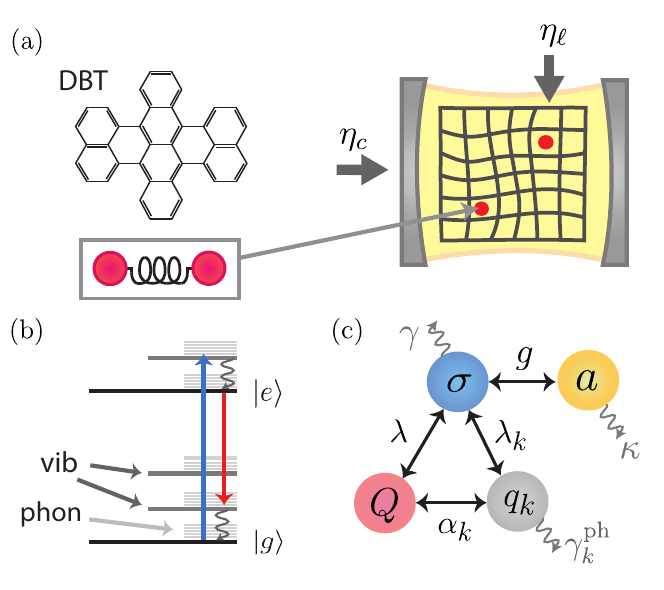}
\caption{\emph{Model}. (a) Schematic representation of a host crystal containing molecular impuritites (e.g.~DBT) which is placed inside an optical cavity. For simplicity we restrict our treatment to a single nuclear coordinate. The molecules are either illuminated directly by  a laser with amplitude $\eta_\ell$ or indirectly driven via the cavity mode with amplitude $\eta_c$.  (b) Jablonski scheme of a molecule in a solid-state environment showing vibrational and phononic sublevels of both electronic excited $\ket{e}$ and ground $\ket{g}$ state. Excitation of the molecule is typically followed by quick vibrational relaxation (wavy lines). (c) Illustration of the relevant couplings between electronic $\{\sigma\}$, localized vibrational $\{Q\}$, phononic $\{q_k\}$ and optical $\{a\}$ degrees of freedom as well as decay processes indicated by wavy lines (spontaneous emission rate $\gamma$, cavity decay rate $\kappa$, and phonon decay rate $\gamma_k^\text{ph}$).}
\label{fig1}
\end{figure}
\indent We develop here a complex model where all interactions between light, electronic transitions, vibrons and phonons are taken into account for finite temperatures. We derive general expressions for the light scattered by a molecular system (of one or more molecules) embedded in a solid-state environment outside or inside an optical cavity [see Fig.~\ref{fig1}(a)]. As schematically illustrated in Fig.~\ref{fig1}(c) the light (mode $a$) couples to electronic transitions (Pauli operator $\sigma$) via a Tavis-Cummings Hamiltonian. These are in turn affected by the vibronic coupling to one or more molecular vibrations which leads to the red-shifted Stokes lines in emission [cf.~Fig.~\ref{fig1}(b)]. We focus here on a single mode with relative motion coordinate $Q$ for the sake of simplicity. The solid-state matrix supports a multitude of bosonic phonon modes with displacements $q_k$ ($k$ from 1 to $N$) which directly modify the electronic transition leading to the occurrence of phonon wings in the emission and absorption spectra. In addition, molecular vibrons can deposit energy into phonons as a displacement-displacement interaction, leading to an irreversible process of vibrational relaxation. \textcolor{black}{We will start with the description of the vibrational relaxation process in Sec.~\ref{vibrationalrelaxation} since all subsequent effects will depend on this mechanism}. We show that linear phonon-vibron couplings can already result in irreversible vibrational relaxation involving both single- and multi-phonon processes. Moreover, such dynamics can be either Markovian or non-Markovian, depending on the relation between the vibrational frequency and the maximum phonon frequency. For closely spaced molecules, the same formalism allows for the derivation of collective relaxation dynamics exhibiting effects similar to super/subradiance in dense quantum emitter systems.  Classical light driving is included in Sec.~\ref{molecularspectroscopy} by calculating absorption spectra for coherently driven molecules under the influence of vibronic and phononic couplings as well as thermal effects. We show that interestingly, the vibronic and electron-phonon couplings do not cause any dephasing dynamics even at high temperatures, i.e.~the zero-phonon line is mainly lifetime-limited in the linear coupling model. Following a quantum Langevin equations approach, we derive absorption spectra for coherently driven molecules under the influence of vibronic and phononic couplings as well as thermal and finite-size effects. Finally, for molecular polaritonic systems in a cavity setting, we derive transmission functions of the cavity field (see Sec.~\ref{cavityspectroscopy}), showing the reduction of the vacuum Rabi splitting with increasing vibronic and phononic coupling, as well as phononic signatures in the Purcell regime. The effect of temperature on the asymmetry of cavity polaritons is quantified by deriving effective rate equations for the polariton cross-talk dynamics.

\subsection{Hamiltonian formulation}
We consider one molecule (later we extend to more than one) embedded in a bulk medium comprised of $N$ unit cells. \textcolor{black}{Our perturbational assumption is that, since the bulk is large, the guest molecule does not significantly change the overall modes of the bulk}. The electronic degrees of freedom of the molecule are denoted by states $\ket{g}$ and $\ket{e}$ with the former at zero energy and the latter at $\omega_0$ (we set $\hbar$ to unity), corresponding to a lowering operator $\sigma=\ket{g}\bra{e}$. We assume only a pair of ground and excited potential landscapes with identical curvature along the nuclear coordinate and make the harmonic approximation, where the motion of the nuclei can be described by a harmonic vibration at frequency $\nu$ and bosonic operators $b$ and $b^\dagger$, satisfying the usual bosonic commutation relations $[b,b^\dagger]=1$.

From the displacement between the minima of the two potential landscapes one obtains a vibronic coupling quantified by a dimensionless factor $\lambda$ (the square root of the Huang-Rhys parameter) and described by a standard Holstein-Hamiltonian \cite{holstein1959study},
\begin{equation}
\label{holsteinelvib}
H_{\text{el-vib}}=-\lambda\nu \sqrt{2}   \sigma^\dagger \sigma Q,
\end{equation}
where $Q=(b+b^\dagger)/\sqrt{2}$ is the dimensionless position operator of the vibronic degree of freedom (the momentum quadrature is given by $P=i(b^\dagger-b)/\sqrt{2}$). The Holstein coupling also leads to a shift of the electronic excited state energy $\omega_0+\lambda^2\nu$, which is removed by the diagonalizing polaron transformation $\mathcal{U}_{\text{el-vib}}$. The polaron transformation $\mathcal{U}_{\text{el-vib}}=e^{i\sqrt{2}\lambda P\sigma^\dagger\sigma}=\ket{g}\bra{g}+\mathcal{B}^\dagger \ket{e}\bra{e}$ can be seen as a conditional displacement affecting only the excited state, where $\mathcal{B}^\dagger=e^{i\sqrt{2}\lambda P}$ is the inverse displacement operator for the molecular vibration creating a coherent state when applied to vacuum: $\mathcal{B}^\dagger\ket{0_\nu}=\ket{-\lambda}$. \textcolor{black}{The Hamiltonian in Eq.~(\ref{holsteinelvib}) does not consider nonadiabatic vibronic coupling which would lead to off-diagonal coupling terms (proportional to $\sigma_x$ and $\sigma_y$) and which could drive electronic transitions. Such nonadiabatic terms become relevant if two potential surfaces come close to each other \cite{ulusoy2019modifying}. In Appendix \ref{offdiagonal} we briefly discuss how one could treat such terms in the Langevin equations of motion.} One could also consider a difference in curvatures between ground (frequency $\nu$) and excited state (frequency $\bar{\nu}$) potential surfaces which would result in a quadratic coupling term $H_{\text{el-vib}}^{\text{quad}}=\beta Q^2\sigma^\dagger\sigma$ with squeezing parameter of the vibrational wavepacket $\beta=(\bar{\nu}^2-\nu^2)/(2\nu)$. We will assume that the vibron \textcolor{black}{quickly} thermalizes with the environment (via the \textcolor{black}{fast} mechanism of vibron-phonon coupling described below) at temperature $T$ and achieves a steady state thermal occupancy $\bar{n}=[\exp({ \nu/(k_\text{B}\text{T})})-1]^{-1}$.

The electronic transition is coupled to the quantum electromagnetic vacuum which opens a radiative decay channel with collapse operator $\sigma$ via spontaneous emission at rate $\gamma$. For a general collapse operator $\cal{O}$ with rate $\gamma_{\cal{O}}$ we model the dissipative dynamics via a Lindblad term ${\cal{L}}_{\cal{O}}[\rho]=\gamma_{\cal{O}}\left\{2\cal{O}\rho \cal{O}^\dagger-\rho\cal{O}^\dagger\cal{O}-\cal{O}^\dagger\cal{O}\rho\right\}$ applied as a superoperator to the density operator $\rho$ of the system. The vibronic coupling leads to the presence of Stokes lines in emission and to a mismatch between the molecular emission and absorption profiles. Following the stochastic quantum evolution of a polaron operator $\tilde{\sigma}=\mathcal{B}^\dagger\sigma$ (vibrationally dressed Pauli operator for the electronic transition) analytical solutions for the absorption and emission spectra of the molecule can be derived in the presence of vibrons~\cite{reitz2019langevin}. \\
\indent In addition to the coupling to internal vibrations of its nuclei, the electronic transition is also modified through coupling to the delocalized phonon modes of the crystal. We describe the bulk modes as a bath of independent harmonic oscillators with bosonic operators $c_k$ and $c_k^\dagger$ and frequencies $\omega_k$. The electron-phonon coupling (see Appendix \ref{derivation} for derivations) can then be cast in the same Holstein form as for the vibron
\begin{equation}
H_{\text{el-phon}}= -\sum_k   \lambda_k \omega_k\sqrt{2}\sigma^\dagger\sigma q_k,
\end{equation}
where the displacement operators refer to each individual collective phonon mode $q_k=(c_k+c^\dagger_k)/\sqrt{2}$ (the momentum operator is given by $p_k=i(c_k^\dagger-c_k)/\sqrt{2}$). The coupling factors $\lambda_k$ depend on the specifics of the molecule and the bulk crystal. Similarly to the vibronic case, the electron-phonon interaction can be diagonalized by means of a polaron transformation $\mathcal{U_{\text{el-phon}}}=\ket{g}\bra{g}+\mathcal{D}^\dagger \ket{e}\bra{e}$, whereby $\mathcal{D}^\dagger=\prod_k \mathcal{D}_k^\dagger=e^{\sum_k i \sqrt{2}\lambda_k p_k}$ is the product of all phonon mode displacements, signifying a collective transformation for all phonon modes. We will assume that the bulk is kept at a constant temperature and is always in thermal equilibrium with the individual mode thermal average occupancies amounting to $\bar{n}_k=[\exp({ \omega_k/(k_\text{B}T)})-1]^{-1}$. The coupling to the phonons gives rise to a multitude of sidebands in the absorption and emission spectra which coalesce into a phonon wing that becomes especially important at elevated temperatures. We will then follow the temporal dynamics of a collective polaron operator $\tilde{\sigma}=\mathcal{D}^\dagger\mathcal{B}^\dagger\sigma$ which includes both vibronic and electron-phonon couplings.\\
\indent Phonons also affect the dynamics of the vibrational mode. Modifications of the bond length associated with the molecular vibration leads to a force on the surrounding crystal (and vice versa), giving rise to a displacement-displacement coupling,
\begin{equation}
\label{Hvibphon}
H_{\text{vib-phon}}=-\sum_k \alpha_k q_k Q\,.
\end{equation}
The coupling coefficients $\alpha_k$ are explicitly derived in Appendix \ref{derivation}. In the limit of large bulk media, this Hamiltonian can lead to an effective irreversible dynamics, i.e.~a vibrational relaxation effect. This is the Caldeira-Leggett model widely treated in the literature as it leads to a non-trivial master equation evolution which cannot be expressed in Lindblad form and is \textcolor{black}{cumbersome} to solve analytically \cite{caldeira1983path, caldeira1981influence, hu1992quantum}. To circumvent this difficulty, we follow the formalism of Langevin equations under the concrete conditions imposed by the one-dimensional situation considered here. We are then in a position to identify the Markovian versus non-Markovian regimes of vibrational relaxation conditioned on the phonon spectrum, namely on the maximum phonon frequency $\omegamax$ of the system. We can additionally account for a finite phonon lifetime by including a  decay rate $\gamma_k^{\text{ph}}$ for each phonon mode. \\
\indent To perform spectroscopy, we add a laser drive modeled as $H_{\ell}=i\eta_{\ell} \left(\sigma^\dagger e^{-i\omega_{\ell}t}-\sigma e^{i\omega_{\ell}t}\right)$ with amplitude $\eta_{\ell}$. \textcolor{black}{We will assume weak driving such that the assumption of thermal equilibrium is still valid}. Furthermore, to treat various aspects of molecular polaritonics, we describe the dynamics of a hybrid light-matter platform by adding the coupling of a confined optical mode at frequency $\omega_c$ to the electronic transition via a Jaynes-Cummings interaction
\begin{equation}
H_\text{JC}= g(a\sigma^\dagger+\sigma a^\dagger).
\end{equation}
The bosonic operator $a$ satisfies the commutation relation $[a,a^\dagger]=1$ and the coupling is given by $g=[d_{\text{eg}}^2\omega_c/(2\epsilon_0 V)]^{1/2}$, where $d_{\text{eg}}$ is the electronic transition dipole moment and $V$ is the quantization volume ($\epsilon_0$ is vacuum permittivity). Spectroscopy of the cavity-molecule system can be done by adding a cavity pump $H_{\ell}=i\eta_{\text{c}} \left(a^\dagger e^{-i\omega_{\ell}t}-a e^{i\omega_{\ell}t}\right)$ with amplitude $\eta_{\text{c}}$. The cavity loss is modeled as a Lindblad process with collapse operator $a$ and rate $\kappa$. In standard cavity QED, depending on the magnitude of the coherent exchange rate $g$ to the loss terms $\kappa$ and $\gamma$ one can progressively advance from a strong cooperativity Purcell regime to a strong coupling regime where polaritons emerge. We will mainly focus on analytical derivations of the effects of electron-vibron and electron-phonon couplings at finite temperatures on the emergence of a spectral splitting in the strong coupling regime as well as the transmission in the Purcell regime.

\section{Vibrational relaxation}
\label{vibrationalrelaxation}

\begin{figure*}[t]
\includegraphics[width=2.0\columnwidth]{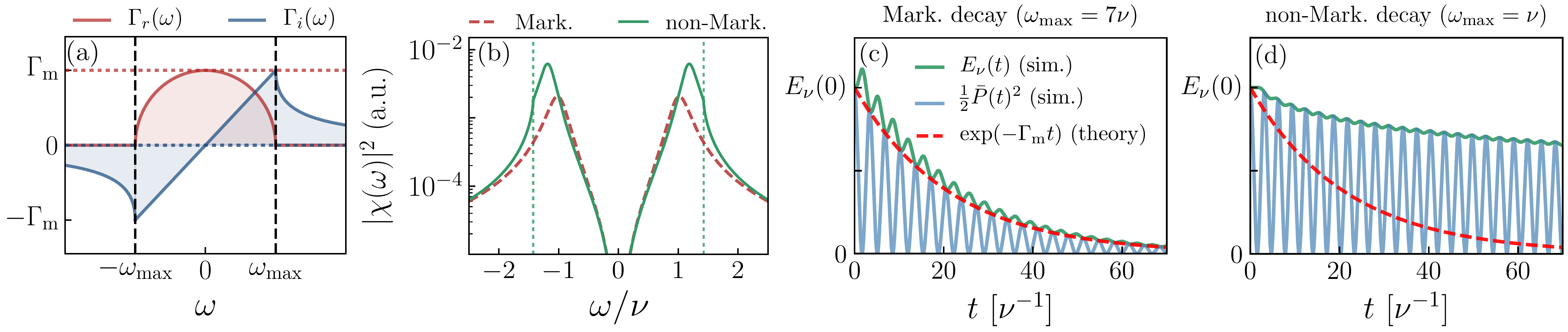}
\caption{\emph{Markovian vs.~non-Markovian vibrational relaxation}. (a)  Real and imaginary part of the frequency-dependent non-Markovian response function $\Gamma(\omega)$. The dotted horizontal lines $\Gamma_r(\omega) =\Gamma_{\text{m}}$ and $\Gamma_i (\omega)=0$ show the Markovian limit $\omegamax\to \infty$. (b) Logarithmic plot of the mechanical susceptibility $|\chi(\omega)|^2$ in the Markovian regime $\omegamax\gg\nu$ (dashed curve) and for a finite cutoff $\omegamax=1.4\nu$ (solid curve). In the latter case, the cutoff at $\pm\omegamax$ is indicated by the vertical dotted lines. (c) Relaxation of molecular vibrational energy $E_\nu (t)$ due to coupling to $N=500$ phonon modes (simulation of classical equations of motion) in the Markovian limit $\omegamax=7\nu$ for $\Gamma_{\text{m}}=\nu/20$ and comparison with Brownian motion theory (red dashed curve). (d) Vibrational relaxation for identical $\Gamma_{\text{m}}$ but in the non-Markovian regime $\omegamax=\nu$. Exponential decay with $\Gamma_{\text{m}}$ (dashed red curve) then fails to predict the behavior. In both cases we assumed a finite phonon lifetime with constant $\mathcal{Q}$-factor $\mathcal{Q}=\omega_k/\gamma_k^\text{ph}=50$.}
\label{fig2}
\end{figure*}

A decay path for the molecular vibration stems from its coupling to the bath of phonon modes supported by the bulk. While it is generally agreed that nonlinear vibron-phonon couplings contribute to the vibrational relaxation process, especially in the higher temperature regime \cite{nitzan2017energy}, we restrict our treatment to a coupling in the bilinear form of Eq.~\eqref{Hvibphon}. To understand the physical picture, we first show that in perturbation theory the bilinear Hamiltonian leads to a competition between fundamental processes that involve the decay of a vibrational quantum into superpositions of either single phonon states or many phonons adding together in energy to the initial vibrational state energy. Afterwards we proceed by writing a set of coupled deterministic equations of motion for the vibrational quadratures of the molecule $\left\{Q,P\right\}$ and the collective normal modes of crystal vibrations $\left\{q_k,p_k\right\}$. This allows for the elimination of the phonon degrees of freedom and the derivation of an effective Brownian noise stochastic evolution model for the molecular vibrations. We illustrate regimes of Markovian and non-Markovian dynamics and show that an equivalent approach tailored to two molecules can lead to collective vibrational relaxation strongly dependent on the molecule-molecule separation.

\subsection{Fundamental vibron-phonon processes}
Let us consider an initial state containing a single vibrational quantum $\ket{1_{\nu}, \text{vac}_{\text{ph}}}$ that evolves according to the vibron-phonon bilinear Hamiltonian of Eq.~\eqref{Hvibphon}. We aim to follow the fundamental processes leading to the energy of the vibration deposited in superpositions of single or multi-phonon states. We move to the interaction picture by removing the free energy with $U = e^{iH_{0}t}$ with the free Hamiltonian $H_{0} = \nu b^{\dagger}b + \sum_{k=1}^{N} \omega_{k} c_{k}^{\dagger}c_{k}$. The time-dependent interaction picture Hamiltonian, thus, becomes
\begin{eqnarray}
\label{Eq2}
\tilde{H}\! =\! -\sum_{k=1}^{N}\! \alpha_{k}\! \left( e^{-i\nu t}b + e^{i\nu t}b^{\dagger}\right) \!\left(e^{-i\omega_{k} t}c_{k} + e^{i\omega_{k} t}c_{k}^{\dagger} \right)\!.
\end{eqnarray}
The formal solution of the Schr\"{o}dinger equation can then be written as a Dyson series $\ket{\phi(t)} = \mathcal{T} e^{-i\int_{0}^{t}d\tau \tilde{H}(\tau)}\ket{1_{\nu}, \text{vac}_{\text{ph}}}$. We can proceed by evaluating the first term in the series which leads to (see Appendix \ref{vibronphonon} for details) resonant scattering ($\omega_{k} = \nu$) into single-phonon states $\ket{0_{\nu},1_k}$ at perturbative rate $\alpha_k t$ as well as off-resonant scattering ($\omega_{k} \neq \nu$) into states $\ket{0_{\nu},1_k}$ with probability inversely proportional to the detuning $\omega_{k} -\nu$. We note that for $\nu > \omega_{\text{max}}$, only off-resonant transitions are possible. The next order of perturbation theory, however, leads to multi-phonon processes where resonant transitions to states containing three phonons $\ket{0_{\nu},1_{j_1},1_{j_2},1_{j_3}}$ become possible. The resonance condition reads $\omega_{j_1} + \omega_{j_2} + \omega_{j_3} =\nu $ for $j_1 \neq j_2 \neq j_3$, and its amplitude is a sum over terms $\alpha_{j_1}\alpha_{j_2}\alpha_{j_3}t/(\omega_{j_2} + \omega_{j_3})(\omega_{j_3}-\nu)$. These terms are small with respect to the rates of the resonances starting in the first order for $\nu \leq \omega_{\text{max}}$ and in total are comparable to the single-phonon scattering off-resonant terms.

\subsection{Effective Brownian noise model}
Formal elimination of the phonon modes (see Appendix \ref{browniansection} for details) leads to an effective Brownian motion equation for the momentum of the vibrational mode
\begin{align}
\label{dotp}
\dot{P}=-\tilde{\nu}Q -\Gamma\ast P+\xi,
\end{align}
while the displacement follows the unmodified equation $\dot{Q}=\nu P$. The effect of the phonon bath is twofold: (i) it can shift the vibrational frequency to $\tilde{\nu}=\nu-\nu_s$ and (ii) it leads to a generally non-Markovian decay kernel expressed as a convolution $\Gamma\ast P=\int_{0}^{\infty}dt'\Gamma(t-t')P(t')$. For the particular case considered in the Appendix, the crystal-induced frequency shift is expressed as
\begin{equation}
\nu_s=\nu\frac{(\Delta k)^2}{2k_0k_{\text{M}}},
\end{equation}
where $k_0$ denotes the spring constant of the host crystal, $k_{\text{M}}$ represents the spring constant of the vibron, and $\Delta k$ is a measure for the coupling of the molecule's relative motion to the bulk. For a discrete system, the expression for the damping kernel $\Gamma(t)=\sum_k \alpha_k^2\nu/\omega_k\cos(\omega_k t)\Theta(t)$ involves a sum over all phonon modes which can be turned into the following expression in the continuum limit ($N\to\infty$)
\begin{equation}
\Gamma(t-t')=\Gamma_{\text{m}}\frac{J_1(\omega_{\text{max}}|t-t'|)}{|t-t'|}\Theta(t-t').
\end{equation}
Here, $J_n(x)$ denotes the $n$-th order Bessel function of the first kind, $\Theta(t)$ stands for the Heaviside function and $\Gamma_{\text{m}}={2\nu\nu_s}/{\omega_{\text{max}}}$ is the decay rate in the Markovian limit. A similar expression is known from the Rubin model \cite{rubin1963momentum}, where one considers the damping of a single mass defect in a 1D harmonic crystal. The zero-average Langevin noise term $\xi$ is determined by the initial conditions of the phonon bath and can be expressed in discrete form as $\xi(t)=\sum_k \alpha_k \left(q_k(0)\cos(\omega_k t)+p_k (0)\sin(\omega_k t)\right)$. We can treat Eq.~(\ref{dotp}) more easily in the Fourier space where the convolution becomes a product
\begin{align}
-i\omega P(\omega)=-\tilde{\nu}Q(\omega)-\Gamma(\omega)P(\omega)+\xi(\omega),
\end{align}
and the Fourier transform of the non-Markovian decay kernel $\Gamma(\omega)$ generally contains a real and imaginary part $\Gamma(\omega)=\Gamma_r(\omega)+i\Gamma_i (\omega)$. Figure~\ref{fig2}(a) shows a plot of $\Gamma_r(\omega)$ and $\Gamma_i(\omega)$ where we can interpret the imaginary part as a frequency shift which is largest around $\omega=\pm\omegamax$. Together with the transformed equation for the position quadrature $-i\omega Q(\omega)=\nu P(\omega)$ we then obtain an algebraic set of equations which allows us to calculate any kind of correlations for the molecular vibration, both in time and frequency domains. This will be needed later on for computing the optical response of the molecule.

\begin{figure*}[t]
\includegraphics[width=2.078\columnwidth]{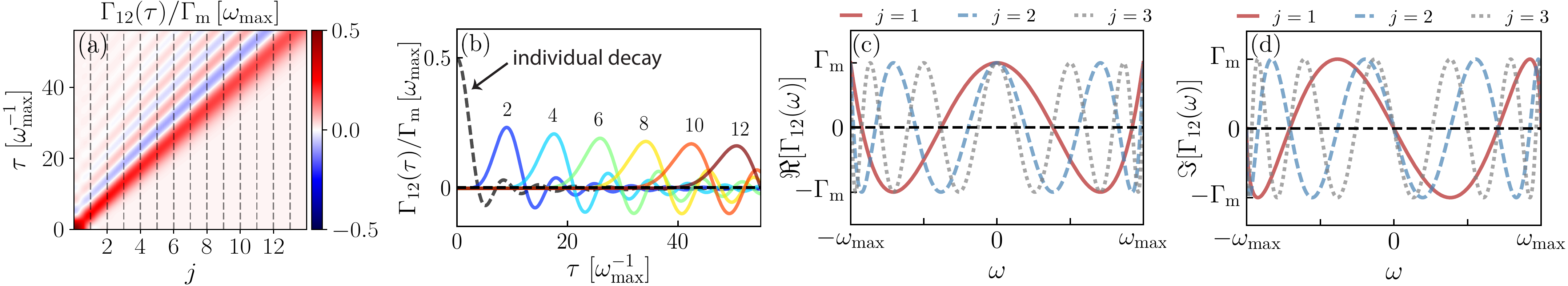}
\caption{\emph{Collective vibrational effects}. (a) Collective interaction kernel $\Gamma_{12}(\tau)=\Gamma_{21}(\tau)$ as a function of time delay $\tau$ and separation $j$. In the crystal, only integer values of $j$ are permitted (dashed vertical lines). (b) Comparison between individual decay (dashed curve, assuming two identical molecules $\Gamma_1(\tau)=\Gamma_2(\tau)$) and collective interaction for increasing distance $j$ (as indicated by the numbers above the curves). (c) Real and (d) imaginary part of $\Gamma_{12}(\omega)$ between $-\omegamax$ and $+\omegamax$ for $j=1$ (solid), $2$ (dashed), and  $3$ (dotted).}
\label{fig3}
\end{figure*}

\subsection{Markovian versus non-Markovian regimes}
The Markovian limit is achieved when the vibrational frequency lies well within the phonon spectrum $\omega_\text{max}\gg\nu$ and $\Gamma(\omega)$ becomes flat in frequency space: $\Gamma(\omega)=\Gamma_{\text{m}}$. In this case the memory kernel tends to a $\delta$-function: $\Gamma(t)=2\Gamma_{\text{m}}\delta(t)\Theta(t)$ with the convention $\Theta(0)=1/2$.  In the continuum limit, the correlations at different times are
\begin{align}
\braket{\xi(t)\xi(t')}=\frac{1}{2\pi}\int_{-\infty}^{\infty}d\omega e^{-i\omega(t-t')} S_{\text{th}}(\omega),
\end{align}
where the noise spectrum is expressed similarly to the case of a standard thermal spectrum for a harmonic oscillator in thermal equilibrium $S_{\text{th}}(\omega)={\Gamma_r(\omega)\omega}/\nu [\coth\left(\beta\omega/2\right)+1]$ in terms of the inverse temperature $\beta=(k_\text{B} T)^{-1}$. The difference lies in the frequency-dependence of the real part of the decay rate function
\begin{align}
 \Gamma_r(\omega)=\Gamma_{\text{m}}\frac{\sqrt{\omega_{\text{max}}^2-\omega^2}}{\omega_{\text{max}}}\Theta(\omega_{\text{max}}\!-\!\omega)\Theta(\omega_{\text{max}}\!+\!\omega),
\end{align}
where the Heaviside functions provide a natural cutoff of the spectrum at $\pm\omega_{\text{max}}$. While in the time domain, the noise is only $\delta$-correlated at high temperatures and $\omega_{\text{max}}\to\infty$, the noise is always $\delta$-correlated (yet colored) in the frequency domain $\braket{\xi(\omega)\xi(\omega')}=S_{\text{th}}(\omega)\delta(\omega+\omega')$. This property is helpful for analytical estimation of the molecular absorption and emission in the presence of non-Markovian vibrational relaxation. In frequency space, the response of the vibron to the input noise of the phonon bath is characterized by the susceptibility $\chi(\omega)={-i\omega}\left[{\nu^2-\omega^2-i\Gamma(\omega)\omega}\right]^{-1}$ defined by $P(\omega)=\chi(\omega)\xi(\omega)$ (for simplicity we assumed $\tilde{\nu}\approx\nu$).

In Fig.~\ref{fig2}(b), we plot $|\chi(\omega)|^2$ for the two cases $\omegamax\gg\nu$ (Markovian limit) and $\omegamax\approx\nu$ (non-Markovian regime) for identical $\Gamma_{\text{m}}$. While in the Markovian regime, the susceptibility has two approximately Lorentzian sidebands with linewidth $\Gamma_{\text{m}}$ centered around $\pm\nu$, the finite frequency cutoff in the non-Markovian case leads to an unconventional lineshape with reduced linewidth and slight frequency shift. In the time domain, we can simulate the microscopic classical equations of motion for a large number of phonon modes and compare the results to the standard Markovian limit obtained from Brownian motion theory. This is illustrated in Figs.~\ref{fig2}(c) and (d) where we simulate the average energy of the vibron mode $E_\nu(t)=\left(\bar{P}(t)^2+\bar{Q}(t)^2\right)/2$ for classical observables $\{\bar{P},\bar{Q}\}$ interacting with $N=500$ phonon modes in the Markovian and non-Markovian regimes, respectively. While one obtains an exponential decay with $\Gamma_{\text{m}}$ in the Markovian regime [Fig.~\ref{fig2}(c)], in the non-Markovian case [Fig.~\ref{fig2}(d)] one finds a slower nonexponential decay (for identical $\Gamma_{\text{m}}$) which does not reach zero for long times.\\
\indent The time domain correlations can be easily computed from the thermal spectrum convoluted with the modified mechanical susceptibility in the Fourier domain
\begin{align}
\label{fouriertransformp}
\braket{P(t)P(t')}=\frac{1}{2\pi}\int_{-\infty}^{\infty}d\omega e^{-i\omega(t-t')}|\chi(\omega)|^2 S_{\text{th}}(\omega),
\end{align}
which also includes the non-Markovian regime. At low temperatures $\beta^{-1}\ll \nu$  the sideband at $-\nu$ is suppressed and the thermal spectrum can be approximated as $S_{\text{th}}(\omega)=[2{\Gamma_r(\omega)\omega}/\nu ] \Theta(\omega)$. This two-time correlation function of the momentum quadrature will be required later in the calculation of molecular spectra in section \ref{molecularspectroscopy}.

\subsection{Collective vibrational effects}

A collection of impurity molecules sitting close to each other within the same crystal will see the same phonon bath and can, therefore, undergo a collective vibrational relaxation process. This is similar to the phenomenon of subradiance/superradiance of quantum emitters commonly coupled to an electromagnetic environment, where the rate of photon emission from the whole system can be smaller or larger than that of an individual isolated emitter. In order to elucidate this aspect, we follow the approach sketched above, i.e.~we eliminate the phonon modes to obtain a set of coupled Langevin equations for two molecules situated $2j$ sites apart from each other:
\begin{subequations}
\label{collectivevib}
\begin{align}
\dot{P}_1=-\tilde{\nu}_1 Q_1-\Omega Q_2-\Gamma_1\ast P_1-\Gamma_{12}\ast P_2+\xi_1,\\
\dot{P}_2=-\tilde{\nu}_2 Q_2 -\Omega Q_1-\Gamma_2\ast P_2-\Gamma_{21}\ast P_1+\xi_2.
\label{collective}
\end{align}
\end{subequations}
The mutually induced (small) energy shift $\Omega=\sum_k {\alpha_{k,1}\alpha_{k,2}}/{\omega_k}$ and the mutual damping kernels $\Gamma_{12}=\nu_2\sum_k {\alpha_{k,1}\alpha_{k,2}}/{\omega_k}\cos(\omega_k t)\Theta(t)$ and $\Gamma_{21}=\nu_1\sum_k {\alpha_{k,1}\alpha_{k,2}}/{\omega_k}\cos(\omega_k t)\Theta(t)$ are strongly dependent on the intermolecular separation $2j$ (see Appendix \ref{collectivevibrationalrelaxation} for full expression), whereas the individual decay terms $\Gamma_1$ and $\Gamma_2$ are given by the expressions derived previously. Importantly, now also the noise terms  $\xi_1$ and $\xi_2$ are not independent of each other but correlated according to a separation-dependent expression specified in Appendix \ref{collectivevibrationalrelaxation}. In the continuum limit $N\to\infty$, the collective interaction kernels can be approximated with the aid of higher-order Bessel functions (assuming identical molecules $\nu_1=\nu_2$ and consequently $\Gamma_{12}(t)=\Gamma_{21}(t)$):
\begin{align}
\Gamma_{12}(t-t')=\Gamma_{\text{m}}\frac{4j J_{4j}(\omegamax |t-t'|)}{|t-t'|}\Theta(t-t').
\end{align}
In Fig.~\ref{fig3}(a), we plot the collective decay kernel as a function of time and intermolecular separations $2j$. The collective effects do not occur instantaneously but in a highly time-delayed fashion [cf.~Fig~\ref{fig3}(b)]. We can interpret the collective interaction as an exchange of phonon wavepackets between the two molecules, where the wavepackets are traveling with the group velocity $v_g =\partial\omega/\partial q$ of the crystal (lattice constant $a$) at a maximum speed $v_g^{\text{max}}\approx a\omegamax/2$ (the high frequency components towards the band edge are slower). This leads to an approximate time of $\tau=4j\omegamax^{-1}$ for the wavepacket to propagate from one molecule to the other.\\
\indent The collective interaction will also lead to a modification of the vibrational lifetimes of the molecules which we want to describe in the following. To this end, one can again proceed with a Fourier analysis of Eqs.~(\ref{collectivevib}). The expression for the non-Markovian collective interaction kernel in frequency space (between $-\omegamax$ and $+\omegamax$) reads
\begin{align}
\Gamma_{12}(\omega) =& i\Gammam U_{4j-1}\left(\frac{\omega}{\omegamax}\right)\frac{\sqrt{\omegamax^2-\omega^2}}{\omegamax}\\\nonumber
&+\Gammam T_{4j}\left(\frac{\omega}{\omegamax}\right),
\end{align}
where we introduced the Chebychev polynomials of first ($T_n $) and second kind ($U_n$). We are interested in the real part of the above expression which will give rise to a collectively-induced modification of the vibrational lifetime while the imaginary part corresponds again to a frequency shift. In Figs.~\ref{fig3}(c) and (d) we plot the real and imaginary parts of $\Gamma_{12}(\omega)$ for small distances $j$, respectively.

In the Markovian limit $\omegamax\gg\nu$ everything becomes flat in frequency space and one can approximate $\Gamma_{12}(\omega)=\Gamma_{12}(0)=\Gamma_{\text{m}}$ and consequently $(\Gamma_{12}\ast P_i)\approx \Gamma_{\text{m}}P_i$ with $i=\{1,2\}$. A diagonalization can be performed by moving into collective quadratures $P_+=P_1+P_2$ and $P_- = P_1 - P_2$ (and identically for the positions) for which the equations of motion decouple
\begin{subequations}
\begin{align}
\dot{P}_+&\approx -(\tilde{\nu}+\Omega)Q_+-2\Gamma_{\text{m}}P_++\xi_1+\xi_2,\\
\dot{P}_-&\approx -(\tilde{\nu}-\Omega)Q_-+\xi_1-\xi_2.
\end{align}
\end{subequations}
While one of the collective modes undergoes relaxation at an increased rate $2\Gamma_{\text{m}}$, the orthogonal collective mode can be eventually decoupled from the phononic environment. Of course, as the derivation we have performed is restricted to one-dimensional crystals, it would be interesting to explore this effect in three dimensional scenarios where both longitudinal and transverse phonon modes have to be considered with effects stemming from the molecular orientation as well as the influence of anharmonic potentials. A recent theoretical work also discusses phonon-bath mediated interactions between two molecular impurities immersed in nanodroplets with respect to the rotational degrees of freedom of the molecules \cite{li2020intermolecular}.

\section{Fundamental spectral features}
\label{molecularspectroscopy}
Let us now consider a molecule driven by a coherent light field. We will make use of and extend the formalism used in Ref.~\cite{reitz2019langevin} to compare the effect of Markovian versus non-Markovian vibrational relaxation, phonon imprint on spectra and temperature effects. To derive the absorption profile of a laser-driven molecule, one can compute the steady-state excited state population $\mathcal{P}_{\text{e}}=\braket{\sigma^\dagger\sigma}=\eta_\ell \left[\braket{\sigma}+\braket{\sigma}^*\right]/(2\gamma)$. The average steady-state dipole moment can formally be written as (\textcolor{black}{note that we are assuming weak driving conditions $\eta_\ell\ll\gamma$ such that the laser drive only probes the linear response of the dipole}):
 \begin{align}
\braket{\sigma}=\eta_\ell\int_{-\infty}^t dt' e^{-\left[\gamma-i\left(\omega_\ell-{\omega}_0\right)\right](t-t')} \braket{\mathcal{B}(t)\mathcal{B}^\dagger (t')}.
\label{integralsigma}
\end{align}
The important quantity to be estimated is the correlation function for the displacement operators of the molecular vibration $\braket{\mathcal{B}(t)\mathcal{B}^\dagger (t')}$, which is fully characterized by the Huang-Rhys factor $\lambda^2$ and the second-order momentum correlation functions:
\begin{equation}
\braket{\mathcal{B}(t)\mathcal{B}^\dagger (t')}=e^{-2\lambda^2\left(\braket{P^2}-\braket{P(t)P(t')}\right)}.
\end{equation}
The stationary correlation $\braket{P(t)^2}=\braket{P^2}=1/2+\bar{n}$ includes the temperature of the environment and does not depend on the details of the decay process. The two-time correlations $\braket{P(t)P(t')}$ (and consequently the vibrational linewidths of the resulting optical spectrum) are crucially determined by the details of the dissipation model \textcolor{black}{derived in section \ref{vibrationalrelaxation}}. In order to capture the non-Markovian character of the vibrational relaxation, we extend the method used in Ref.~\cite{reitz2019langevin} by computing correlations in the Fourier domain and then transforming to the time domain.

\subsection{The non-Markovian vibrational relaxation regime}

Let us first consider the imprint of the particularities of the vibrational relaxation process onto the absorption and emission spectra when molecule-light interactions are taken into account.
 For the calculation of the momentum correlation function $\braket{P(t)P(t')}$ one has to evaluate  the integral in Eq.~(\ref{fouriertransformp}), where the susceptibility weighted with the thermal spectrum is given by the general expression
\begin{figure}[t]
\includegraphics[width=1.02\columnwidth]{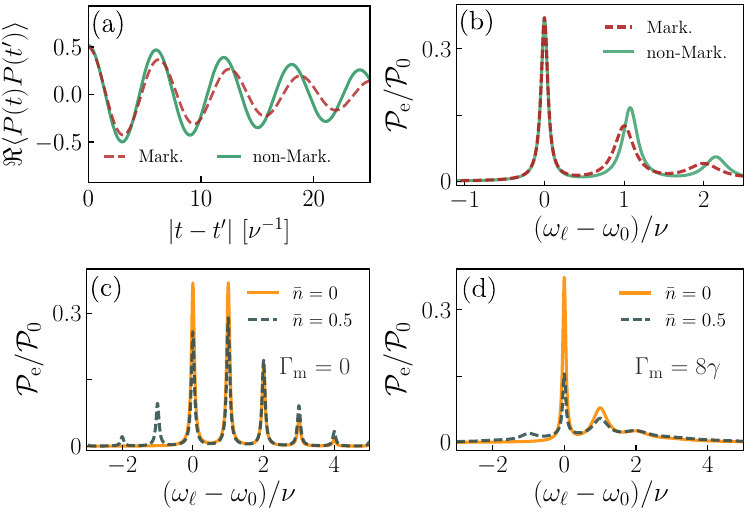}
\caption{\emph{Molecular spectroscopy}. (a) Real part of the correlation function $\braket{P(t)P(t')}$ for Markovian decay (dashed) versus non-Markovian decay (solid, cutoff at $\omegamax=1.3\nu$) for $\Gamma_{\text{m}} =0.1\nu$ and at zero temperature $\bar{n}=0$. (b) Comparison between the resulting absorption spectra (steady state population $\mathcal{P}_{\text{e}}$ normalized by steady-state population of resonantly driven two-level system $\mathcal{P}_0$) in the Markovian and non-Markovian regimes for the same parameters as in (a) and $\lambda=1$, $\gamma=\Gamma_{\text{m}}/4$. (c) Effect of thermal occupation  $\bar{n}$ on absorption spectra ($\lambda=1$) without vibrational relaxation $\Gamma_{\text{m}}=0$ and (d) including vibrational relaxation $\Gamma_{\text{m}}=8\gamma$ (assuming Markovian decay).}
\label{fig4}
\end{figure}
\begin{align}
\label{nonmarksusc}
|\chi(\omega)|^2 S_{\text{th}}(\omega)=\frac{\Gamma_r (\omega)\omega^3\left[\coth(\frac{\beta\omega}{2})+1\right]/\nu}{\left(\nu^2\! -\! \omega^2\! +\Gamma_i (\omega)\omega \right)^2\! +\! \Gamma_r(\omega)^2\omega^2},
\end{align}
with $\Gamma_i(\omega)=\Gamma_{\text{m}} {\omega}/\omegamax$ between $-\omegamax$ and $+\omegamax$. As discussed in the previous section, the real part of $\Gamma(\omega)$ determines the decay rate while the imaginary part leads to a frequency shift. Generally, performing the integral over the expression in  Eq.~(\ref{nonmarksusc}) is difficult since the line shapes can be very far from simple Lorentzians. However, assuming a good oscillator ($\Gamma_{\text{m}} \ll \nu$) and consequently a sharply peaked susceptibility that only picks frequencies around the vibrational resonance,  we can obtain an effective modified frequency $\nu'$ and decay rate $\Gamma'$ in the non-Markovian regime (however assuming $\omegamax>\nu$) with $\nu'=\left[\nu^2+\Gamma_i(\nu)\nu\right]^{1/2}$ and $\Gamma'=\Gamma_r(\nu')$. By expanding Eq.~(\ref{nonmarksusc}) around the poles of the denominator $\omega=\pm\nu'+\delta$ and assuming $|\delta|\ll|\nu'|$, one can then calculate the temperature-dependent momentum correlation function in the non-Markovian regime:
\begin{align}
\braket{P(t)P(t')}\! =\! \left[\!\left(\bar{n}\!+\!\frac{1}{2}\right)\!\cos(\nu'\tau)\!-\!\frac{i}{2}\sin(\nu'\tau)\!\right]\! e^{-\frac{\Gamma'}{2}|\tau|},
\end{align}
with time delay $\tau=t-t'$. This allows for an analytical evaluation of the integral in Eq.~(\ref{integralsigma}) (see Appendix \ref{calcabsorptionspectrum} for detailed calculation) and leads to a steady-state excited-state population of
\begin{align}
\label{pss}
\frac{\mathcal{P}_{\text{e}}}{\eta_\ell^2}=\!\sum_{n=0}^\infty\sum_{l=0}^n\frac{L(n)B(n,l)\left(\gamma\!+\!n\frac{\Gamma'}{2}\right)/\gamma}{(\gamma\!+\!n\frac{\Gamma'}{2})^2+\left[(\omega_\ell\!-\!\omega_0)-(n\!-\!2l){\nu'}\right]^2},
\end{align}
where we introduced $L(n)=e^{-\lambda^2(1+2\bar{n})}\frac{\lambda^{2n}}{n!}$ and $B(n,l)=\binom{n}{l}\left(\bar{n}+1\right)^{n-l}\bar{n}^l$ . One can immediately obtain the result for the Markovian limit by replacing $\nu'\to\nu$ and $\Gamma'\to\Gamma_{\text{m}}$. Figures \ref{fig4}(a),(b) show a comparison between the momentum correlation function and the resulting steady-state population (normalized by the steady-state population of a resonantly driven two-level system $\mathcal{P}_0={\eta_\ell^2}/{\gamma^2}$) in the Markovian and non-Markovian regimes (for fixed $\Gamma_{\text{m}}$). We can see that non-Markovianity leads to modified spectral positions and linewidths of the vibronic sidebands while the ZPL is not affected by the dissipation process. The denominator of Eq.~(\ref{pss}) contains a sum over Lorentzians with a series of blue-shifted lines with index $n$ arising from the electron-vibration interaction which are weighted by a Poissonian distribution. Thermal occupation of the vibrational states can counteract this effect however by leading to red-shifted lines in absorption [see Figure \ref{fig4}(c)] with index $l$ and weighted by a binomial distribution. However, as shown in Figure \ref{fig4}(d), for large vibrational relaxation rates $\Gamma_{\text{m}}\gg\gamma$ the sidebands will be suppressed and absorption and emission of the molecule will mostly occur on the ZPL transitions $\ket{g,m_\nu}\leftrightarrow\ket{e,m_\nu}$. While in the case of zero temperature the ZPL is solely determined by $n=0$, for finite temperatures all terms with $n=2l$ can contribute to it.

An important quantity is the Franck-Condon factor $f_{\text{FC}}$ which measures the reduction of the ZPL intensity due to coupling to internal vibrations. This factor is given by the average displacement squared $f_{\text{FC}}=\braket{\mathcal{B}^\dagger}^2=e^{-\lambda^2(1+2\bar{n})}$ and does not depend on the vibrational relaxation of the molecule.
 Using the fact that $(\bar{n}+1)/\bar{n}=e^{\beta\nu}$, one can express Eq.~(\ref{pss}) as a sum over just a single index in the limit $2\lambda^2\sqrt{\bar{n}(\bar{n}+1)}\ll 1$ (see Appendix \ref{calcabsorptionspectrum} for derivation) as
\begin{align}
\frac{\mathcal{P}_{\text{e}}}{\eta_\ell^2}=\sum_{n=-\infty}^\infty \frac{f_{\text{FC}}\left(\frac{\bar{n}+1}{\bar{n}}\right)^{\frac{n}{2}}I_n\left(2\lambda^2\bar{N}\right)\left(\gamma\!+\!|n|\frac{\Gamma'}{2}\right)/\gamma}{(\gamma+|n|\frac{\Gamma'}{2})^2+(\omega_\ell-\omega_0-n\nu')^2},
\end{align}
 with $I_n (x)$ denoting the \textit{modified} Bessel functions of the first kind and $\bar{N}=\sqrt{\bar{n}(\bar{n}+1)}$. This expression is similar to the result known from the standard Huang-Rhys theory for emission and absorption \cite{huang1950theory}, but it now additionally includes vibrational relaxations $\Gamma'$. The ZPL contribution ($n=0$) at resonance is thus simply given by $\mathcal{P}_{\text{e}}={\eta_\ell^2f_{\text{FC}}}/{\gamma^2}$. The emission spectrum can be calculated from the Fourier transform of two-time correlations $\braket{\sigma^\dagger (\tau)\sigma (0)}$. Considering the decay of an initially excited molecule, one finds that the emission spectrum is simply given as the mirror image (with respect to the ZPL) of the absorption spectrum which is why we restrict ourselves to the calculation of the absorption profile.
\begin{figure*}[t]
\includegraphics[width=1.96\columnwidth]{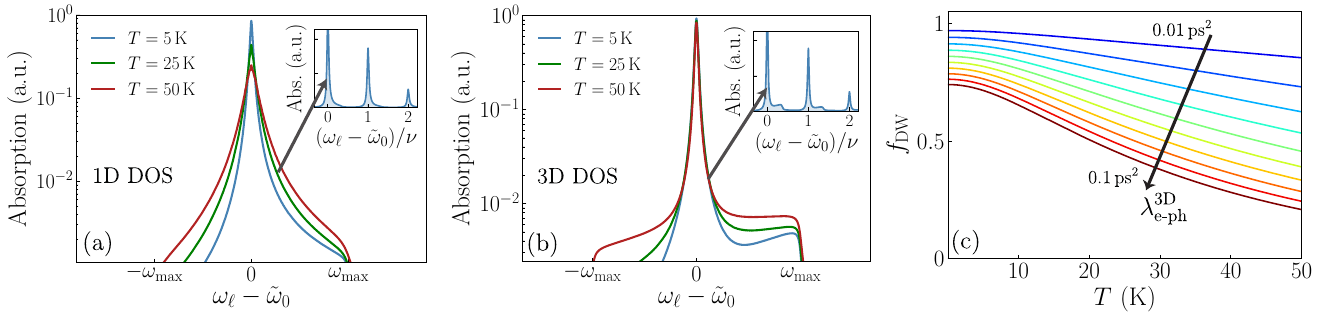}
\caption{\emph{Phonon imprint on absorption}. Absorption spectra (logarithmic scale) of zero-phonon line including phonon wing at different temperatures for (a) 1D density of states and $\lambda_{\text{e-ph}}^{\text{1D}}=0.03$ and (b) 3D density of states and  $\lambda^{\text{3D}}_{\text{e-ph}}=0.02\,\mathrm{ps^2}$ with $\omegamax=3\,\mathrm{THz}$ in both cases. Insets show schematic of total molecular absorption spectrum with vibrational sidebands accompanied by phonon wings. (c) Debye-Waller factor $f_{\text{DW}}$ for a 3D density of states as a function of temperature with increasing coupling strength $\lambda_{\text{e-ph}}^{\text{3D}}$ (as indicated by arrow) in equidistant steps from $\lambda_{\text{e-ph}}^{\text{3D}}=0.01\,\mathrm{ps^2}$ to $\lambda_{\text{e-ph}}^{\text{3D}}=0.1\,\mathrm{ps^2}$ and cutoff frequency of $\omegamax=3\,\mathrm{THz}$.}
\label{fig5}
\end{figure*}

\subsection{Phonon imprint on spectra}

So far, we have considered the phonons only as a bath that can provide vibrational relaxation for the molecule and have neglected the effect of electron-phonon coupling. However, this can become a dominant mechanism at larger temperatures where all acoustic and optical phonon modes are thermally activated ($> 50\,\text{K}$) and the probability of a ZPL transition is very small. To include electron-phonon coupling, the expression for the steady-state dipole moment [cf.~Eq.~(\ref{integralsigma})] has to also account for the displacement of the electronic excited state caused by the phonons
\begin{align}
\label{ephsigma}
 \braket{\sigma}&=\\\nonumber
&\eta_\ell\!\int_{-\infty}^t\! dt' e^{-\left[\gamma-i\left(\omega_\ell-\tilde{{\omega}}_0\right)\right](t-t')}\! \braket{\mathcal{B}(t)\mathcal{B}^\dagger (t')}\braket{\mathcal{D}(t)\mathcal{D}^\dagger (t')}.
\end{align}
Here, the coupling to phonons additionally leads to a renormalization of the electronic transition frequency $\tilde{\omega}_0=\omega_0-\sum_k \lambda_k^2\omega_k$ (polaron shift). \textcolor{black}{The expression in Eq.~(\ref{ephsigma}) now jointly contains all of the effects: electron-phonon coupling, electron-vibron coupling and vibrational relaxation (through the correlation function $\braket{\mathcal{B}(t)\mathcal{B}^\dagger(t')}$)}. Since we consider phonon modes to be independent of each other, the displacement correlation function \textcolor{black}{of the phonons} can be factorized $\braket{\mathcal{D}(t)\mathcal{D}^\dagger(t')}=\prod_k\braket{\mathcal{D}_k(t)\mathcal{D}_k^\dagger(t')}$ where the correlation for each single mode is given by $\braket{\mathcal{D}_k(t)\mathcal{D}_k^\dagger(t')}=\mathrm{exp}\left[-{2\lambda_k^2 \left(\braket{p_k^2}-\braket{p_k(t)p_k(t')}\right)}\right]$. When replacing the sum over $k$ with an integral over $\omega$ in the continuum limit, this yields (neglecting phonon decay as it will not influence the spectra in the continuum limit):
\begin{align}
\label{correlationphonons}
\braket{\mathcal{D}(t)\mathcal{D}^\dagger (t')}&=\\\nonumber
&e^{\int_0^{\infty}\!d\omega\! \frac{J(\omega)}{\omega^2}\left[\coth\left(\frac{\beta\omega}{2}\right)\left(\cos\left(\omega\tau\right)-1\right)-i\sin\left(\omega\tau\right)\right]}.
\end{align}
Here, we have introduced the spectral density of the electron-phonon coupling $J(\omega)=\sum_k |\lambda_k\omega_k|^2\delta(\omega-\omega_k) =n(\omega)\lambda(\omega)^2\omega^2$ where $n(\omega)$ denotes the density of states. In the one-dimensional derivation considered here we obtain for the spectral density
\begin{align}
J^{\text{1D}}(\omega)=\lambda_{\text{e-ph}}^{\text{1D}}\cdot\omega\frac{\sqrt{\omegamax^2-\omega^2}}{\omegamax}\Theta(\omegamax-\omega).
\end{align}
The electron-phonon coupling constant $\lambda_{\text{e-ph}}^{\text{1D}}$ is derived in Appendix \ref{elphcoupling} and depends, among other things, on the displacement of the crystal atoms upon excitation of the molecule as well as on the spring constants between the molecule's atoms and the neighboring crystal atoms. Again, the cutoff at $\omegamax$ arises naturally from the dispersion of the crystal. In the continuum limit considered here, this spectral density would lead to a divergence of the integral in Eq.~(\ref{correlationphonons}) due to the high density of low-frequency phonons, which is a well known problem for 1D crystals \cite{kikas1996anomalous, hizhnyakov2012zero}. This issue can be addressed by considering  only a finite-sized 1D crystal with a minimum phonon frequency cutoff $\omega_{\text{min}}>0$. However, one can instead also consider a spectral density stemming from a 3D density of states:
\begin{align}
J^{\text{3D}}(\omega)=\lambda_{\text{e-ph}}^{\text{3D}}\cdot\omega^3\frac{\sqrt{\omegamax^2-\omega^2}}{\omegamax}\Theta(\omegamax-\omega),
\end{align}
where the electron-phonon coupling constant $\lambda_{\text{e-ph}}^{\text{3D}}$ now has units $[\text{s}^2]$.
In Figs.~\ref{fig5}(a) and (b) we plot the resulting absorption spectrum of the ZPL for 1D and 3D densities of states, whereby the exact shape of the phonon wing is determined by the spectral density function $J(\omega)$. While analytical expressions for the integral in Eq.~(\ref{correlationphonons}) are difficult to obtain in the continuum case, we can express the absorption spectrum of the ZPL including phonon sideband in terms of discrete lines as
\begin{align}
\frac{\mathcal{P}_{\text{e}}}{\eta_\ell^2}\!=\!\sum_{\{n_k\}}^{\infty}\!\sum_{\{l_k\}}^{\{n_k\}}\frac{\prod_{k=1}^{N}L_k(n_k)B_k(n_k,l_k)}{\gamma^2\!+\!\left[\omega_\ell-\tilde{\omega}_0\!-\!\sum_{k=1}^N (n_k\!-\!2l_k)\omega_k\right]^2}.
\end{align}
Here the sum runs over all $\{n_k\}=n_1,\hdots,n_N$ and $\{l_k\}=l_1,\hdots,l_N$. This can be seen as a generalization of the result in Eq.~(\ref{pss}) for many modes where the $N$ phonon modes are indexed by $k$ and the function $L_k (n_k)$ accounts for the displacement of the excited state while the binomial distribution $B_k (n_k, l_k)$ accounts for the thermal occupation of each mode. As one can see in Figs.~\ref{fig5}(a) and (b), thermal occupation of the phonons leads to red-shifted phonon sidebands in absorption and eventually to a symmetric absorption spectrum around the zero-phonon line in the limit of large temperatures. Note that here we did not explicitly include phonon decay $\gamma_k^{\text{ph}}$ as it does not influence the absorption spectra in the continuum limit (the phonon peaks overlap and are not resolved). However, one can easily account for a finite phonon lifetime by including it in the momentum correlations $\braket{p_k (t)p_k (t')}=\frac{1}{2}[(1\!+\!2\bar{n}_k)\cos(\omega_k \tau)\!-\!i\sin(\omega_k \tau)]e^{-\gamma_k^{\text{ph}}|\tau|}$.
 Similarly to the Franck-Condon factor for vibrons, one defines the Debye-Waller factor $f_\text{DW}=\braket{\mathcal{D}^\dagger}^2=\mathrm{exp}\left[{-\int_0^\infty d\omega  J(\omega)\omega^{-2}\coth(\beta\omega/2) }\right]$ which measures the reduction of the ZPL intensity due to the scattering of light into phonons. In Fig.~\ref{fig5}(c) we show the behavior of the $f_\text{DW}$ in the 3D case for different coupling strengths at low temperatures, revealing a stronger temperature-dependence for larger couplings. The total reduction of the ZPL intensity as compared to the two-level system case is then given by the product $f_{\text{FC}}\cdot f_{\text{DW}}$.
\subsection{Dephasing}
Within the model we consider, where all interactions stem from a harmonic treatment of both intramolecular vibrations and crystal motion, the zero-phonon linewidth of the electronic transitions is largely independent of temperature. In reality, to account for higher temperature effects one needs contributions quadratic in the phononic displacements which has been theoretically pointed out and experimentally observed \cite{muljarov2004dephasing,jakubczyk2016impact}. However, even in the linear regime, the fact that vibronic and electron-phonon couplings do not lead to significant dephasing is a non-trivial result. One could e.g.~expect that the Holstein-Hamiltonian for electron-phonon coupling $H_{\text{H}}=\left[\omega_0-\sum_k \sqrt{2}\lambda_k\omega_k q_k\right]\sigma^\dagger \sigma+\sum_k \omega_k c_k^\dagger c_k$ which sees a stochastic shift of the excited electronic level should lead to a dephasing of the ground-excited coherence $\braket{\sigma}$. One reason is the similarity to the pure dephasing of a two-level transition subjected to a noisy laser undergoing evolution with the Hamiltonian $[\omega_0+\dot{\phi}(t)]\sigma^\dagger \sigma$ where the frequency is continuously shaken by a white noise stochastic term of zero average and obeying $\braket{\dot{\phi}(t)\dot{\phi}(t')}=\gamma_{\text{deph}} \delta(t-t')$. It is straightfoward to show that the time evolution of the coherence in this case becomes $\braket{\sigma(t)}=\braket{\sigma(0)}e^{-i\omega_0 t}e^{-\gamma_\text{deph}t}$ such that the correlations of the noise indicate the increase in the linewidth of the transition \cite{plankensteiner2016laser}. Similarly, one could expect that the zero-averaged quantum noise stemming from the shaking of the electronic transition in the Holstein-Hamiltonian would lead to the same kind of effect. However, computing the exact time evolution of the coherence in the interaction picture [with Hamiltonian $\tilde{H}_{\text{H}}(t)$] one obtains:
\begin{align}
 \braket{\sigma(t)}& = \braket{\sigma(0)} \mathcal{T}\{ e^{-i\int_0^t dt' \tilde{H}_{\text{H}}(t')} \}e^{-\gamma t} \\\nonumber
&= \braket{\sigma(0)} e^{-(\gamma+i\tilde{\omega}_0) t}\braket{\mathcal{D}(t)\mathcal{D}^\dagger(0)},
\end{align}
where the time-ordered integral can be resolved by a second-order Magnus expansion, confirming the result already known from the polaron picture. The correlation $\braket{\mathcal{D}(t)\mathcal{D}^\dagger(0)}=e^{-\lambda_k^2(2\bar{n}_k+1)}e^{\varphi(t)}$ with $\varphi(t)=\lambda_k^2\left[(2\bar{n}_k+1)\cos(\omega_k t)-i\sin(\omega_k t)\right]$ (for a single mode $\omega_k$)  shows a cosine-term similar to the dephasing but which does not continuously increase in time. For small times $t\ll\omega_k^{-1}$, the cosine-term can be expanded and the dephasing rate can be approximated by $\gamma_{\text{deph}}=\lambda_k^2(\bar{n}_k+1/2)\omega_k^2 t$ while for larger times the rate goes to zero (the time scale is set by $\gamma^{-1}$). In the continuum limit, the time-dependent dephasing rate $\gamma_{\text{deph}}(t)=-\Re\left[\dot{\varphi} (t)\right]$ expresses as
\begin{align}
\gamma_{\text{deph}}(t)=\int_0^\infty d\omega \frac{J(\omega)}{\omega}\coth\left(\frac{\beta\omega}{2}\right)\sin(\omega t).
\end{align}
 In accordance with Figs.~\ref{fig5}(a),(b) we can see that linear Holstein coupling can consequently lead to a temperature-dependent zero-phonon line if there is a large density of low frequency (long wavelength) phonon modes with $\omega_k< \gamma$ which is the case in 1D but not in higher dimensions. This peculiarity of dephasing in 1D has already been discussed in the literature \cite{reichman1996on, kikas1996anomalous, hizhnyakov2012zero}. It is however also well established within the literature that the major contribution of the experimentally observed temperature-dependent broadening of the zero-phonon line is caused by a higher-order electron-phonon interaction of the form \cite{muljarov2004dephasing, osadko1985dephasing, reichman1996on, reigue2017probing}
\begin{align}
H_{\text{el-phon}}^{\text{quad}}=\sum_{k, k'}\beta_{k k'} q_k q_{k'} \sigma^\dagger\sigma,
\end{align}
with the coupling constant of the quadratic interaction $\beta_{k k'}$. This form of the interaction can stem either, within the harmonic assumption, from a difference in curvatures between the ground and excited state potential surfaces or from anharmonic potentials.

\section{Molecular Polaritonics}
\label{cavityspectroscopy}

It is currently of great interest to investigate the behavior of hybrid platforms containing organic molecules interacting with confined light modes such as provided by optical cavities \cite{walther2006cavity,haroche1989cavity,wang2017coherent} or plasmonic nanostructures \cite{chikkaraddy2016singlemolecule, zengin2015realizing}. Such light-dressed platforms have been studied both at the single- and few-molecule level \cite{wang2017coherent, wang2019turning, chikkaraddy2016singlemolecule} as well as in the mesoscopic, collective strong-coupling limit \cite{shalabney2015coherent, lidzey1999room, holmes2004strong}. In these cases, the strong light-matter coupling leads to the formation of polaritonic hybrid states with both light and matter components. Experimental and theoretical works are currently exploring fascinating enhanced properties such as exciton and charge transport \cite{orgiu2015conductivity,schachenmayer2015cavity,feist2015extraordinary,hagenmuller2017cavity,hagenmuller2018cavity}, superconductive behavior \cite{sentef2018cavity, thomas2019exploring} and modified chemical reactivity \cite{hutchinson2012modifying,galego2016suppressing,herrera2016cavity,martinezmartinez2018can,kampschulte2018cavity}. \textcolor{black}{There is also recent interest in the modification of nonadiabatic light-matter dynamics at so-called conical intersections leading to fast nonradiative decay of electronic excited states \cite{ulusoy2019modifying, vendrell2018collective}}. It has been recently shown that the Purcell regime of cavity QED can result in a strong modification of the branching ratio of a single molecule and suppress undesired Stokes lines~\cite{wang2019turning}. Recent theoretical works account for the vibronic coupling of molecules by solving a Holstein-Tavis-Cummings Hamiltonian which leads to the occurence of polaron-polariton states, i.e.~light-matter states where the hybridized states between the bare electronic transition and the light field additionally get dressed by the vibrations of the molecules \cite{herrera2018theory, zeb2018exact, herrera2017dark, neuman2018origin, wu2016when, litinskaya2004fast, kansanen2019theory,reitz2019langevin}. Many models rely on numerical simulations and are based on following the evolution of state vectors under simplified assumptions assuming only vibronic interactions and finite temperature effects. We employ here the approach of the last section and add a Jaynes-Cummings interaction of a molecule in the phononic environment with a localized cavity mode. A weak laser drive maps the intracavity molecular polaritonics effects to the cavity transmission profile, identifying polariton cross-talk effects at any temperature. Furthermore, we map the combined effect of vibronic and electron-phonon interactions onto the cavity output field.

\subsection{Cavity transmission}

\begin{figure*}[t]
\includegraphics[width=1.99\columnwidth]{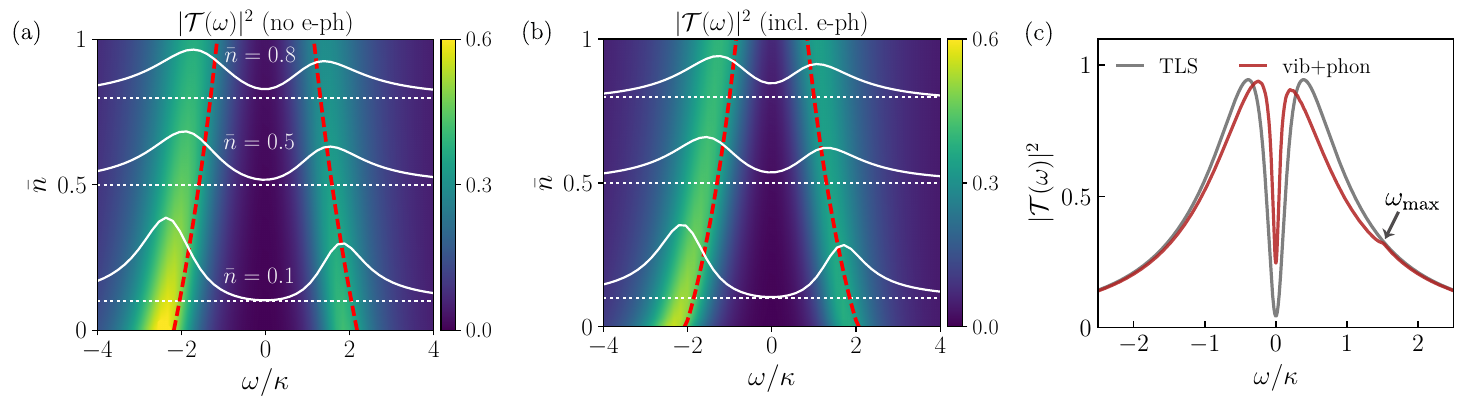}
\caption{\emph{Cavity-molecule spectroscopy}. Normalized cavity transmission $|\mathcal{T}(\omega)|^2$ at resonance $\omega_c=\omega_\ell$ in strong coupling as a function of frequency and thermal occupation $\bar{n}$  for (a) neglecting electron-phonon coupling (b) including electron-phonon coupling (using 3D density of states) for $\lambda=0.8$, $\lambda_{\text{e-ph}}^{\text{3D}}=0.03\,\text{ps}^2$. The white lines show cross sections of the transmission profile for different thermal occupations $\bar{n}$. The dashed red lines show the effective Rabi splitting $2g_{\text{eff}}$. Other parameters: $\nu=6\,\text{THz}$, $\omegamax=3\,\text{THz}$, $g=\nu/2, \Gamma_{\text{m}}=0.08\nu$, $\kappa=1\,\text{THz}$. (c) Comparison of transmission in the Purcell regime between a pure two-level system (TLS) and a molecule subject to electron-phonon and electron-vibron coupling for $\lambda=0.8$, $\lambda_{\text{e-ph}}^{\text{3D}}=0.2\,\text{ps}^2$ at a temperature of $T=10\,\mathrm{K}$. Other parameters: $\omegamax=3\,\mathrm{THz}=1.5\kappa$, $\nu=6\,\mathrm{THz}=3\kappa$, $g=0.35\kappa$.}
\label{fig6}
\end{figure*}

We will consider a cavity mode which is driven with amplitude $\eta_c$ and start with a set of coupled Lagevin equations for the electric field operator $a$ as well as the polaron operator $\tilde{\sigma}(t)=\mathcal{D}^\dagger(t)\mathcal{B}^\dagger(t)\sigma(t)$ in a rotating frame at the laser frequency $\omega_\ell$:
\begin{subequations}
\begin{align}
\label{equationscavitya}
\dot{a}&=-[\kappa-i(\omega_\ell-\omega_c)]a-ig\sigma+\sqrt{2\kappa}A_{\text{in}},\\
\dot{\tilde{\sigma}}&=-[\gamma\!-\!i(\omega_\ell\!-\!\tilde\omega_0)]\tilde{\sigma}\!-\!i g \mathcal{D}^\dagger\mathcal{B}^\dagger a +\!\sqrt{2\gamma}\mathcal{D}^\dagger\mathcal{B}^\dagger{\sigma}_{\text{in}},
\label{equationscavityb}
\end{align}
\end{subequations}
where we defined the effective cavity input $A_{\text{in}}=\eta_c/\sqrt{2\kappa}+a_{\text{in}}$ with zero-average input noise $a_{\text{in}}$ but non-vanishing correlation $\braket{a_{\text{in}}(t)a_{\text{in}}^\dagger(t')}=\delta(t-t')$. The electronic transition is also affected by a white noise input $\sigma_{\text{in}}$ with non-zero correlation $\braket{\sigma_{\text{in}}(t)\sigma_{\text{in}}^\dagger(t')}=\delta(t-t')$. We can formally integrate Eq.~(\ref{equationscavityb}) and substitute it in Eq.~(\ref{equationscavitya}):
\begin{widetext}
\begin{equation}
\braket{\dot{a}}=-\left[\kappa-i(\omega_\ell-\omega_c)\right]\braket{a}-g^2\int_{-\infty}^\infty dt' e^{-[\gamma-i(\omega_\ell-\tilde\omega_0)](t-t')} \Theta(t-t') \braket{\mathcal{D}(t)\mathcal{D}^\dagger (t')}\braket{ \mathcal{B}(t) \mathcal{B}^\dagger(t')} \braket{a(t')}+\eta_c\,,
\label{cavitysubs}
\end{equation}
\end{widetext}
where we took the averages and assume factorizability between optical and vibronic/phononic degrees of freedom which is valid if the timescales of vibrational relaxation and cavity decay are separated, e.g.~$\Gamma_{\text{m}}\gg \kappa$. We notice that the second term in Eq.~(\ref{cavitysubs}) represents a convolution since the correlation functions $\braket{\mathcal{D}(t)\mathcal{D}^\dagger (t')}$ and $\braket{\mathcal{B}(t)\mathcal{B}^\dagger (t')}$ only depend on the time difference $|t-t'|$. Denoting $\mathcal H(t-t')=e^{-[\gamma-i(\omega_\ell-\tilde\omega_0)](t-t')}\Theta(t-t') \braket{\mathcal{D}(t)\mathcal{D}^\dagger (t')}\braket{ \mathcal{B}(t) \mathcal{B}^\dagger(t')}$, the normalized cavity transmission amplitude  $\mathcal{T}(\omega)=\frac{\braket{A_{\text{out}}(\omega)}}{\braket{A_{\text{in}}(\omega)}}$ can be derived from input-output relations as
\begin{align}
\mathcal{T}(\omega)=\frac{\kappa}{g^2 \mathcal H(\omega)-i\omega+\left[\kappa-i(\omega_\ell-\omega_c)\right]},
\end{align}
where $\mathcal H(\omega)$ is the Fourier transform of $\mathcal H(t)$ and describes the optical response of the molecule to the light field \textcolor{black}{including electron-phonon, electron-vibron and vibron-phonon coupling}. If we neglect electron-phonon interactions ($\lambda_{\text{e-ph}}=0$) and assume, for the sake of simplicity, Markovian decay for the vibration (this can also be extended to the non-Markovian regime,  \textcolor{black}{see Section (\ref{molecularspectroscopy})}), $H(\omega)$ acquires the form
\begin{align}
H(\omega)=\sum_{n=0}^\infty \sum_{l=0}^n \frac{L(n)B(n,l)}{(\gamma+n\frac{\Gamma_{\text{m}}}{2})\!-\!i[({\omega\!+\!\omega_\ell\!-\! \omega_0})-({n\! -\! 2l})\nu]}.
\end{align}
Again, the above expression indicates a series of sidebands with strength determined by the Huang-Rhys factor $\lambda^2$ and dependent on the thermal occupation $\bar{n}$. In the case of large \textcolor{black}{vibrational relaxation} $\Gamma_{\text{m}}\gg \gamma$ (corresponds to typical experimental situation), however, those sidebands are suppressed and the cavity will mostly couple to the ZPL transition. We can then define an effective Rabi coupling for the ZPL
\begin{align}
\label{effective}
g_{\text{eff}}=g\sqrt{ f_{\text{FC}}\cdot f_{\text{DW}}}\,,
\end{align}
which takes into account the reduction of the oscillator strength due to Franck-Condon and Debye-Waller factors. In Figs.~\ref{fig6}(a) and (b) we plot the cavity transmission at resonance $\omega_c=\omega_\ell$ for increasing thermal occupation with and without the influence of phonons and find that the splitting of the polariton modes is well described by Eq.~(\ref{effective}). This also manifests itself in the transmission signal in the Purcell regime characterized by weak coupling $g\!<\!|\kappa-\gamma|/2$, but large cooperativity $C=g^2/(\kappa\gamma)\gg 1$ which is a more realistic regime in currently available single-molecule experiments \cite{wang2017coherent, wang2019turning}. In Figure \ref{fig6}(c) we compare the transmission of a pure two-level system (obtained by setting $\lambda=0$, $\lambda_{\text{e-ph}}=0$) with a molecule in a solid-state environment. Here the ZPL appears as a dip in the transmission profile with an increase in width $\tilde{\gamma}=\gamma(1+C_{\text{eff}})$ proportional to the effective cooperativity $C_{\text{eff}}=g_{\text{eff}}^2/(\kappa\gamma)$. As compared to the two-level system case, the coupling to vibrons and phonons leads to a reduction in both width and depth of the antiresonance. If the cavity bandwidth is comparable to the maximum phonon frequency $\omegamax$, the imprint of the phonon wing can also be detected in the transmission signal of the cavity, which is relevant for plasmonic scenarios characterized by large bandwidths \cite{chikkaraddy2016singlemolecule} [see Fig.~\ref{fig6}(c)]. The sidebands of vibrons typically lie at frequencies outside the bandwidth of the cavity $\nu\gg\kappa$ and are consequently unmodified.

\subsection{Vibrationally mediated polariton cross-talk}

As shown in the previous sections, vibronic and electron-phonon couplings reduce the oscillator strength of the molecule and lead to decoherence and are consequently considered as detrimental. However such couplings can also lead to interesting new physics: In Ref.~\cite{reitz2019langevin} it was already shown that vibrations can couple upper and lower polaritonic states in a dissipative fashion, resulting in an effective transfer of population from upper to lower polariton and consequently an asymmetric cavity transmission profile with a suppressed upper polaritonic peak (at zero temperature $\bar{n}=0$) and dominant emission occuring from the lower polariton (this can also be seen in Figs.~\ref{fig6}(a) and (b)). We derive here a more general expression for the population transfer between polaritons showing that for finite thermal occupations of the vibrational mode $\bar{n}$ also a transfer from lower to upper polariton can be activated. Diagonalizing the Jaynes-Cummings part of the Hamiltonian at resonance $\omega_c=\omega_0$ by introducing annihilation operators for upper and lower polariton, $U=(a+\sigma)/\sqrt{2}$ and $L=(a-\sigma)/{\sqrt{2}}$, the Holstein part of the Hamiltonian [Eq.~(\ref{holsteinelvib})] gives rise to a vibration-mediated interaction between upper and lower polariton
\begin{align}
H_{\text{int}}=\frac{\lambda\nu}{2}(U^\dagger L+L^\dagger U)(b^\dagger+b).
\end{align}
\begin{subequations}
This can be interpreted as an exchange interaction which is mediated by either the destruction or creation of a vibrational quantum. From this one can derive equations of motion for the populations of upper and lower polaritonic states:
\begin{align}
\dot{\mathcal{P}}_U&=-2\gamma_+ \mathcal{P}_U +\lambda\nu\Im\braket{U^\dagger L (b^\dagger+b)},\\
\dot{\mathcal{P}}_L&=-2\gamma_- \mathcal{P}_L +\lambda\nu\Im\braket{L^\dagger U (b^\dagger+b)},
\end{align}
\end{subequations}
with the hybridized decay rates of upper and lower polaritonic state $\gamma_\pm = (\kappa+\gamma)/2$ and one can see that the term $\Im\braket{U^\dagger L (b^\dagger+b)}$ is the one responsible for population transfer between the polaritons. In the limit of fast textcolor{blue}{vibrational relaxation} $\Gamma_{\text{m}}\gg\kappa$ this can be turned into a set of rate equations with an effective excitatation transfer from the upper polariton to the lower polariton $\kappa_+$ and a transfer from the lower polariton to the upper polariton $\kappa_-$ (for detailed calculation see Appendix \ref{polcrosstalk}):
\begin{subequations}
\begin{align}
\dot{\mathcal{P}}_U&=-(2\gamma_+ +\kappa_+)\mathcal{P}_U +\kappa_-\mathcal{ P}_L,\\
\dot{\mathcal{P}}_L&=-(2\gamma_-+\kappa_-) \mathcal{P}_L +\kappa_+\mathcal{P}_U.
\end{align}
\end{subequations}
Under the assumption of weak vibronic coupling as compared to the splitting between upper and lower polaritonic state $\lambda\nu\ll 2g$, the rates can be calculated to first order as (again we assume Markovian decay for the vibration for that sake of simplicity):
\begin{subequations}
\begin{align}
\label{polaritonrates}
\kappa_{+}=\frac{1}{4}\frac{\lambda^2\nu^2\Gamma_{\text{m}}(\bar{n}+1)}{(\Gamma_{\text{m}}/2)^2+\left(\omega_+-\omega_--\nu\right)^2},\\
\kappa_{-}=\frac{1}{4}\frac{\lambda^2\nu^2\Gamma_{\text{m}}\bar{n}}{(\Gamma_{\text{m}}/2)^2+\left(\omega_+-\omega_--\nu\right)^2}.
\end{align}
\end{subequations}
Energy transfer between the polaritons can consequently occur if the Rabi splitting $\omega_+-\omega_-\approx 2g$ is roughly equal to the vibrational frequency. In the case of zero temperature ($\bar{n}=0$) the above equations reduces to the results presented in \cite{reitz2019langevin} using a Lindblad decay model for the vibration instead of a Brownian noise model. The ratio $\kappa_-/\kappa_+=\bar{n}/(\bar{n}+1)$ which can be inferred from the polariton heights (for normalized Lorentzians the height and width are connected) and which tends to unity in the limit $\bar{n}\gg 1$ can be seen as a direct measure for temperature as it does not depend on any other parameters. While for single molecules the condition $\omega_+-\omega_-\approx \nu$ is difficult to achieve for vibrational modes in the THz range, this can be achieved in the collective strong coupling regime for many molecules where the coupling grows as $g\sqrt{N}$ or for single molecules with phononic modes in the GHz regime. \textcolor{black}{We also note that, in a similar fashion to the linear coupling, also quadratic electron-phonon and vibronic coupling gives rise to a vibrationally-mediated polariton cross-talk with coupling $H_{\text{int}}^{\text{quad}}=\beta(U^\dagger L+L^\dagger U)Q^2/2$ (for a single vibrational mode). To this end, one could again derive effective rate equations for the quadratically-mediated population transfer between the polaritons in a similar fashion as for the linear coupling case.}

\section{Discussions. Conclusions}
\indent We have provided a new approach based on quantum Langevin equations for the analysis of the fundamental quantum states of molecules and their coupling to their surroundings. These features, which lie at the heart of molecular polaritonics, go well beyond the electronic degrees of freedom and address phenomena such as electron-vibron and electron-phonon couplings as well as vibron-phonon interactions resulting in the relaxation of molecular vibrations. In particular, we have provided analytical expressions for spectroscopic quantities such as molecular absorption and emission inside and outside optical cavities in the presence of vibrations and phonons at any temperature. Moreover, we have presented a model of vibrational relaxation that takes into account the structure of the surrounding phonon bath and makes a distinction between Markovian and non-Markovian regimes. We have demonstrated that the vibrational relaxation of a molecule is crucially determined by the structure of the bath, especially by the maximum phonon frequency $\omega_{\text{max}}$. For two molecules embedded in the same crystalline environment, we have shown that the vibrational modes of the spatially separated molecules can interact with each other, resulting in collective dissipative processes that allow for weaker relaxation of collective vibrations. In the strong coupling regime of cavity QED, we have derived temperature-dependent transfer rates for vibrationally mediated cross-talk between upper and lower polaritonic states, i.e.~hybrid light-matter states that are normally uncoupled in cavity QED studies of atomic systems.  In this work, we based our model on first-principle derivations of the relevant coupling strengths between a single nuclear coordinate of a molecule embedded in a 1D chain. However, the calculations could be readily extended to 3D scenarios and compared with ab-initio calculations for real materials. We point out that our theory could also be relevant for vacancy centers in diamond where similar interactions between electronic degrees of freedom and both localized and delocalized phonon modes occur \cite{albrecht2013coupling, londero2018vibrational}. In the future we want to address the influence of higher-order interactions such as quadratic electron-phonon and vibron-phonon couplings, which are known to play an important role at elevated temperatures. \textcolor{black}{It could also be interesting to consider the cavity-modification of the nonradiative relaxation of molecules at conical intersections \cite{ulusoy2019modifying}}.  We also plan to investigate the collective radiation states of dense molecular ensembles in confined electromagnetic environments such as e.g.~occuring in organic semiconductor microcavities.

\textit{Note added.} Recently, we became aware of a related study \cite{clear2020phonon}. \\

\section{Acknowledgments}
We acknowledge financial support from the Max Planck Society and from the German Federal Ministry of Education and Research, co-funded by the European Commission (project RouTe), project number 13N14839 within the research program "Photonik Forschung Deutschland" (C.~S, V.~S.~and C.~G.).

\bibliography{PhononsRef1}
\onecolumngrid
\newpage
\appendix

\section{Derivation of coupling terms}
\label{derivation}

\begin{figure}[t]
\center
\includegraphics[width=0.65\columnwidth]{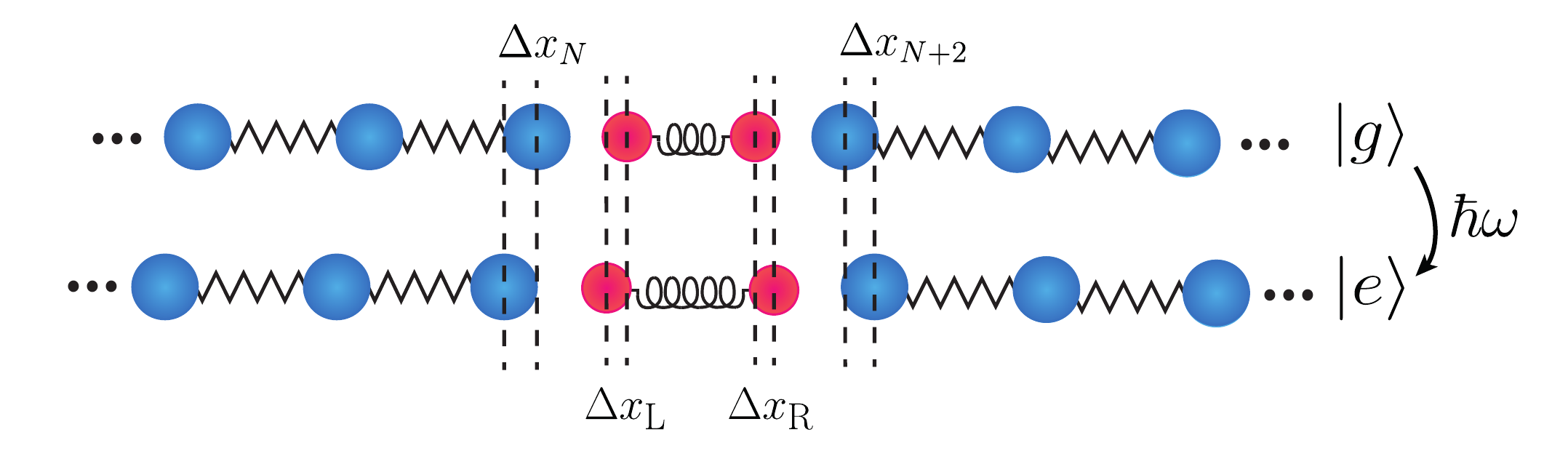}
\caption{\emph{Derivation of coupling terms}. Electronic excitation of a diatomic molecule will lead to a modified bond length and consequently to a push of the surrounding host crystal. }
\label{SMDerivation}
\end{figure}

We want to derive the coupling terms for electron-phonon, electron-vibron and vibron-phonon coupling. To this end, we consider a diatomic molecule (single vibrational mode) with reduced mass $\mu=\frac{m_{\text{L}} m_{\text{R}}}{m_{\text{L}} +m_{\text{R}}}$ which is embedded in a 1D chain with lattice constant $a$ [see Figure \ref{SMDerivation}] and $2N+1$ unit cells (we assume the molecule to be at position $N+1$). Excitation of the molecule will reconfigure the electronic orbitals and bond length of the molecule and will consequently lead to a modified potential between the molecule and the crystal atoms. Therefore, the equilibrium positions of the lattice will depend on the electronic state of the molecule and will be shifted in the excited state by $\Delta x_\ell = x_\ell^e- x_\ell^g$. This interaction can be described by the general Hamiltonian
\begin{align}
H_{\text{bulk-mol}}=\sum_{\ell}\left[\sigma\sigma^\dagger\left(W_{\text{L}\ell}^g+W_{\text{R}\ell}^g\right)+\sigma^\dagger\sigma\left(W_{\text{L}\ell}^e +W_{\text{R}\ell}^e\right)\right],
\end{align}
where $W_{\text{L}\ell}^{g/e}$ and $W_{\text{R}\ell}^{g/e}$ are the potentials in ground and excited state between the left and right nuclei of the dopant and a lattice site $\ell$. Assuming harmonic potentials between the guest and host atoms, we can expand around the equilibrium positions in the excited state (assuming that the vibrational frequencies in ground and excited state are identical)
\begin{subequations}
\label{expansion}
\begin{align}
W_{\text{L}\ell}^e (|x_{\text{L}}-x_\ell|) &= \hbar\omega_{0}+\frac{1}{2}k_{\text{L}\ell}(x_{\text{L}}-x_{\ell}-\Delta x_{\text{L}\ell})^2,\\
W_{\text{R}\ell}^e (|x_{\text{R}}-x_\ell|)&=\hbar\omega_{0}+\frac{1}{2}k_{\text{R}\ell}(x_{\text{R}}-x_{\ell}-\Delta x_{\text{R}\ell})^2,
\end{align}
\end{subequations}
where $\Delta x_{\text{L}\ell}=\Delta x_{\text{L}}-\Delta x_\ell$ and $\Delta x_{\text{R}\ell}=\Delta x_{\text{R}}-\Delta x_\ell$ account for the displaced equilibrium positions in the excited state and the spring constants $k_{\text{L}\ell}$, $k_{\text{R}\ell}$ are given as the second derivatives of the potentials evaluated at the excited state equilibrium positions.
Kinetic and potential energy of the bare molecule is described by
\begin{align}
H_{\text{mol}}=\frac{p_{\text{L}}^2}{2m_{\text{L}}}+\frac{p_{\text{R}}^2}{2m_{\text{R}}}+V^g(|x_{\text{R}}-x_{\text{L}}|) \sigma\sigma^\dagger + V^e(|x_{\text{R}}-x_{\text{L}}|) \sigma^\dagger\sigma,
\end{align}
and Taylor expansion leads to
\begin{align}
V^e (|x_{\text{R}}-x_{\text{L}}|) = \hbar\omega_{0}+\frac{1}{2}k_{\text{M}}\left(x_{\text{R}}-x_{\text{L}}-\Delta x\right)^2,
\end{align}
with $\Delta x =\Delta x_{\text{R}}-\Delta x_{\text{L}}$. For the molecular two-body system it is convenient to go into center-of-mass and relative motion coordinates with $x_{\text{cm}}=\frac{m_{\text{L}}x_{\text{L}}+m_{\text{R}}x_{\text{R}}}{m_{\text{L}}+m_{\text{R}}}$, $x_{\text{rm}}=x_{\text{R}}-x_{\text{L}}$. Center-of-mass and relative-motion momenta are given by
\begin{subequations}
\begin{align}
p_{\text{cm}}&=p_{\text{L}}+p_{\text{R}},\\
p_{\text{rm}}&=\mu\left(v_{\text{R}}-v_{\text{L}}\right)=\frac{m_{\text{L}}p_{\text{R}}-m_{\text{R}}p_{\text{L}}}{m_{\text{L}}+m_{\text{R}}},
\end{align}
\end{subequations}
and the inverse transformations are given as  $p_{\text{R}}=\frac{m_{\text{R}}}{m_{\text{R}}+m_{\text{L}}}p_{\text{cm}}+p_{\text{rm}}$ and $p_{\text{L}}=\frac{m_{\text{L}}}{m_{\text{R}}+m_{\text{L}}}p_{\text{cm}}-p_{\text{rm}}$. With this we obtain for the molecule term (defining $M=m_{\text{L}}+m_{\text{R}}$):
\begin{equation}
\label{molecularhamiltonian}
H_{\text{mol}}=\hbar\omega_0\sigma^\dagger\sigma+\frac{p_{\text{cm}}^2}{2M}+\frac{p_{\text{rm}}^2}{2\mu}+\frac{1}{2}k_{\text{M}} x_{\text{rm}}^2+\left(\frac{1}{2}k_{\text{M}}\Delta x^2-k_{\text{M}} x_{\text{rm}}\Delta x\right)\sigma^\dagger\sigma,
\end{equation}
from which we see that an electronic excitation couples to the relative motion of the nuclei:
\begin{align}
H_{\text{el-vib}}=-k_{\text{M}}x_{\text{rm}}\Delta x \sigma^\dagger\sigma,
\end{align}
and leads to additional energy shift of $\frac{1}{2}k_{\text{M}}\Delta x^2\sigma^\dagger\sigma$.
 We can now re-express the potentials from Eqs.~(\ref{expansion}) in terms of center-of-mass and relative motion of the molecule (defining $W_{\text{L}\ell}=\sigma\sigma^\dagger W_{\text{L}\ell}^g+\sigma^\dagger\sigma W_{\text{L}\ell}^e$, $W_{\text{R}\ell}=\sigma\sigma^\dagger W_{\text{R}\ell}^g+\sigma^\dagger\sigma W_{\text{R}\ell}^e$):
\begin{subequations}
\begin{align}
W_{\text{L}\ell}&=\frac{1}{2}k_{\text{L}\ell}\left(x_{\text{cm}}-\frac{m_{\text{R}}}{M}x_{\text{rm}}-x_\ell\right)^2-k_{\text{L}\ell}\left(x_{\text{cm}}-\frac{m_{\text{R}}}{M}x_{\text{rm}}-x_\ell\right)\Delta x_{\text{L}\ell}\sigma^\dagger\sigma+\frac{1}{2}k_{\text{L}\ell}\Delta x_{\text{L}\ell}^2\sigma^\dagger\sigma,\\
W_{\text{R}\ell}&=\frac{1}{2}k_{\text{R}\ell}\left(x_{\text{cm}}+\frac{m_{\text{L}}}{M}x_{\text{rm}}-x_\ell\right)^2-k_{\text{R}\ell}\left(x_{\text{cm}}+\frac{m_{\text{L}}}{M}x_{\text{rm}}-x_\ell\right)\Delta x_{\text{R}\ell}\sigma^\dagger\sigma+\frac{1}{2}k_{\text{R}\ell}\Delta x_{\text{R}\ell}^2\sigma^\dagger\sigma.
\end{align}
\end{subequations}
We are mostly interested in the coupling between the phonons and the relative motion of the molecule (which will give rise to vibrational relaxation) as well as the coupling between the electronic excitation and the phonons. In the expanded equations above we can see that the first term couples couples the relative motion of the molecule $x_{\text{rm}}$ to the motion of the lattice atoms $x_\ell$, while the second term couples $x_{\ell}$ to the electronic excitation $\sigma^\dagger\sigma$ and the last term is a constant energy shift. For left and right atom together we then arrive at the interaction Hamiltonians
\begin{subequations}
\label{couplings1}
\begin{align}
H_{\text{vib-phon}}&=-\frac{1}{2}\sum_\ell \left(k_{\text{R}\ell}-k_{\text{L}\ell}\right)x_\ell x_{\text{rm}},\\
H_{\text{el-phon}}&=-\sum_\ell  \Delta x_\ell \left(k_{\text{R}\ell}+k_{\text{L}\ell}\right)x_\ell\sigma^\dagger\sigma,
\end{align}
\end{subequations}
where for simplicity we assumed equal masses $m_{\text{L}}=m_{\text{R}}$ and used that for symmetric molecules $\Delta x_{\text{L}}=-\Delta x_{\text{R}}$.

\subsection{Bulk solution}
\begin{figure}[ht]
\center
\includegraphics[width=0.6\columnwidth]{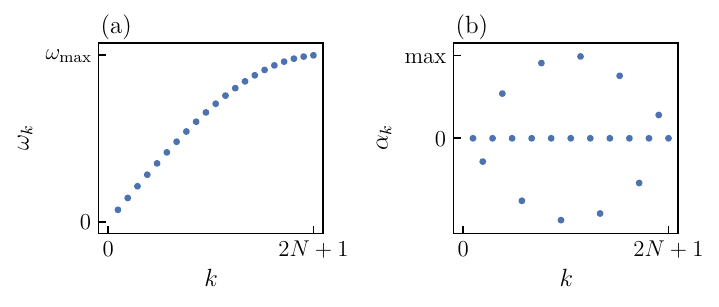}
\caption{\emph{Finite-sized crystal.} a) Dispersion relation $\omega_k$. b) Vibration-phonon coupling $\alpha_k$ (the electron-phonon coupling $\lambda_k$ shows a similar behavior).}
\label{PlotsSM1}
\end{figure}

The contribution of the bulk adds as (just considering nearest-neighbour interactions)
\begin{align}
H_{\text{bulk}}=\sum_\ell\left(\frac{p_\ell^2}{2m}+\frac{1}{2}k_0\left(x_\ell-x_{\ell+1}\right)^2\right),
\end{align}
with the main assumption that the presence of the molecule does not significantly change the bulk modes. Newton's equations of motion for the displacements of the lattice atoms can be arranged in matrix form as
\begin{align}
\ddot{\mathbf{x}}=-\mathbf{M}\mathbf{x},
\end{align}
where $\mathbf{x}=(x_{1},\hdots,x_{2N+1})$ and
\begin{align}
\mathbf{M}=\frac{1}{m}\begin{pmatrix} 2k_0&-k_0 &&0\\
-k_0 & 2k_0&\ddots&\\
&\ddots &\ddots&-k_0 \\
0&&-k_0&2k_0
\end{pmatrix},
\end{align}
where we assume a periodic lattice and neglect the edge modes in our analysis. This is a trigonal Toeplitz matrix which can be easily diagonalized ($\mathbf{D}=\mathbf{T}^{-1}\mathbf{M}\mathbf{T}$) where the $k$-th eigenvalue is given by
\begin{align}
\omega_k^2=\frac{4k_0}{m}\sin^2\left(\frac{\pi k}{2(2N+2)}\right)=\omegamax^2 \sin^2\left(\frac{\pi k}{2(2N+2)}\right),
\end{align}
with $k\in [1,2N+1]$ and the associated normalized eigenvectors read
\begin{align}
\mathbf{v}_{k}=\sqrt{\frac{1}{N+1}}\left(\sin\left(\frac{\pi k}{2N+2}\right),\hdots,\sin\left(\frac{(2N+1)\pi k}{2N+2}\right)\right),
\end{align}
such that in the diagonal basis ($\mathbf{u}=\mathbf{T}^{-1}\mathbf{x}$) the equation of motion for the k-th mode is given by
\begin{align}
\ddot{u}_k=-\omega_k^2 u_k.
\end{align}
We can now express the couplings from Eqs.~(\ref{couplings1}) in terms of the normal modes of the crystal. Further more we will consider only nearest-neighbour coupling  for the molecule, i.e.~$k_{\text{R}\ell},k_{\text{L}\ell}=0$ for $\ell\neq\{N,N+2\}$. We obtain with $x_\ell=\sum_{k=1}^{2N+2}\left(T_{\ell k}\right)u_k$
\begin{subequations}
\label{qs}
\begin{align}
x_N&=\sqrt{\frac{1}{N+1}}\sum_k\sin\left(\frac{\pi k N}{2N+2}\right)u_k,\\
x_{N+2}&=\sqrt{\frac{1}{N+1}}\sum_k\sin\left(\frac{\pi k (N+2)}{2N+2}\right)u_k,
\end{align}
\end{subequations}
and consequently
\begin{subequations}
\label{couplings2}
\begin{align}
H_{\text{vib-phon}}&=-\frac{1}{2}\Delta k \left(x_{N+2}-x_N\right)x_{\text{rm}}=-\Delta k x_{\text{rm}}\sqrt{\frac{1}{N+1}}\sum_k \cos\left(\frac{\pi k }{2}\right)\sin\left(\frac{\pi k }{2N+2}\right) u_k ,\\
H_{\text{el-phon}}&=-\Delta x_{N+2}k_{\text{tot}}\left(x_{N+2}-x_N\right)\sigma^\dagger\sigma=-2\Delta x_{N+2}k_{\text{tot}}\sqrt{\frac{1}{N+1}}\sum_k  \cos\left(\frac{\pi k }{2}\right)\sin\left(\frac{\pi k }{2N+2}\right)u_k\sigma^\dagger\sigma.
\end{align}
\end{subequations}
where we used that $\Delta x_{N+2}=-\Delta x_{N}$ and defined $\Delta k = k_{\text{R}(N+2)}-k_{\text{L}(N+2)}=k_{\Le N}-k_{\R N}$, $k_{\text{tot}}=k_{\R (N+2)}+k_{\Le (N+2)}=k_{\R N}+k_{\Le N}$. We now quantize the Hamiltonians by introducing the position operators (and analog momentum operators) $x_{\text{rm}}=x_{\text{zpm}}(b^\dagger+b)$, $u_k=u_{\text{zpm}}^{(k)}\left(c_k^\dagger+c_k\right)$ with $x_{\text{zpm}}=\sqrt{1/(2\mu\nu)}$, $u_{\text{zpm}}^{(k)}=\sqrt{1/(2m_0\omega_k)}$. We obtain
\begin{subequations}
\label{couplings3}
\begin{align}
H_{\text{el-vib}}&=-\lambda\nu\sqrt{2}Q\sigma^\dagger\sigma,\\
H_{\text{vib-phon}}&=-\sum_k \alpha_k q_k Q ,\\
H_{\text{el-phon}}&=-\sum_k\lambda_k\omega_k\sqrt{2}q_k\sigma^\dagger\sigma,
\end{align}
\end{subequations}
where we have introduced the couplings
\begin{subequations}
\begin{align}
\lambda &=\mu\nu Q_{\text{zpm}}\Delta x,\\
\alpha_k &=2\Delta k \sqrt{\frac{1}{N+1}}\cos\left(\frac{\pi k }{2}\right)\sin\left(\frac{\pi k }{2N+2}\right)u_{\text{zpm}}^{(k)}x_{\text{zpm}},\\
\lambda_k &=2\frac{k_{\text{tot}}}{\omega_k}\sqrt{\frac{1}{N+1}}\cos\left(\frac{\pi k}{2}\right)\sin\left(\frac{\pi k }{2N+2}\right)u_{\text{zpm}}^{(k)}\Delta x_{N+2}.
\label{lambdak}
\end{align}
\end{subequations}
We have further more introduced the dimensionless quadratures $Q=\left(b^\dagger+b\right)/\sqrt{2}$, $P=i\left(b^\dagger-b\right)/\sqrt{2}$ as well as $q_k=\left(c_k^\dagger+c_k\right)/\sqrt{2}$, $p_k=i\left(c_k^\dagger-c_k\right)/\sqrt{2}$ such that the following commutation relations hold: $[Q, P]=i$, $[q_k, p_{k'}]=i\delta_{kk'}$ and the Hamiltonian of the bulk can be expressed as $H_{\text{bulk}}=\sum_k \omega_k\left({p_k^2}+{q_k^2}\right)/2=\sum_k \omega_k c_k^\dagger c_k$. Figure \ref{PlotsSM1} shows a plot of the dispersion $\omega_k$ as well as the vibron-phonon coupling $\alpha_k$ (the electron-vibron coupling shows a similar behavior).

\subsection{Quadratic interaction}

In the previous sections we always assumed identical curvatures of electronic ground and excited states. Here we show that non-identical curvatures give rise to a quadratic electron-vibron interaction (similar for the electron-phonon interaction). Assuming different vibrational frequencies of electronic ground ($\nu$) and excited state ($\bar{\nu}$) the molecular Hamiltonian [Eq.~(\ref{molecularhamiltonian})] becomes:
\begin{align}
H_{\text{mol}}=\omega_0\sigma^\dagger\sigma+\frac{p_{\text{cm}}^2}{2M}+\frac{p_{\text{rm}}^2}{2\mu}+\frac{1}{2}\mu\nu^2 x_{\text{rm}}^2\sigma\sigma^\dagger+\left(\frac{1}{2}\mu\bar{\nu}^2\Delta x^2+\frac{1}{2}\mu\bar{\nu}^2 x_{\text{rm}}^2-\mu\bar{\nu}^2 x_{\text{rm}}\Delta x\right)\sigma^\dagger\sigma.
\end{align}
Quantizing with $x_{\text{rm}}=\sqrt{{1}/({2\mu\nu})}\left(b^\dagger+b\right)$ and $p_{\text{rm}}=i\sqrt{({\mu\nu})/{2}}\left(b^\dagger-b\right)$, this gives rise to an electron-vibron interaction of the form
\begin{align}
H_{\text{el-vib}}=\left[-\lambda\bar{\nu}\sqrt{2}Q+\beta Q^2\right]\sigma^\dagger\sigma,
\end{align}
where the second part accounts for the squeezing of the vibrational wavepacket when going from ground to excited state with coupling strength $\beta=(\bar{\nu}^2-\nu^2)/(2\nu)$.

\section{Vibrational relaxation}
\label{browniansection}

The evolution of the molecule's vibrational mode can be calculated from Hamilton's equations of motion
\begin{subequations}
\begin{align}
\dot{Q}&=\nu P,\\
\dot{P}&=-\nu Q+\sum_k \alpha_k q_k
\label{equationprm}
\end{align}
\end{subequations}
The evolution of the phonon bath degrees of freedom is goverend by
\begin{subequations}
\label{phononbath}
\begin{align}
\dot{q}_k&=\omega_k p_k,\\
\dot{p}_k&=-\omega_k q_k+\alpha_k Q.
\label{equationprm}
\end{align}
\end{subequations}
We can derive an effective equation of motion for $Q$ and $P$ by eliminating the bath degrees of freedom.
We can write Eqs.~(\ref{phononbath}) in matrix form $\dot{\mathbf{v}}=\mathbf{M}\mathbf{v}+\mathbf{v}_{\text{inhom}}$ with $\mathbf{v}=(q_k, p_k)^{\text{T}}$ and $\mathbf{v}_{\text{inhom}}=(0, \alpha_k Q)^{\text{T}}$. The solution is given by
\begin{align}
\mathbf{v}(t)=\mathbf{T}e^{\mathbf{\Lambda}t}\mathbf{T}^{-1}\mathbf{v}(0)+\int_0^t dt' \mathbf{T} e^{\mathbf{\Lambda}(t-t')}\mathbf{T}^{-1}\mathbf{v}_{\text{inhom}}(t'),
\end{align}
with $\mathbf{M}=\mathbf{T}\mathbf{\Lambda}\mathbf{T}^{-1}$, $\mathbf{\Lambda}=\mathrm{diag}(i\omega_k,-i\omega_k)$ and
\begin{align}
   \mathbf{T}=\frac{1}{\sqrt{2}}
  \left[ {\begin{array}{cc}
   -i & i \\
   1 & 1\\
  \end{array} } \right],\qquad
  \mathbf{ T}^{-1}=\frac{1}{\sqrt{2}}
  \left[ {\begin{array}{cc}
   i & 1 \\
   -i & 1\\
  \end{array} } \right].
\end{align}
With this we can derive the solution for $q_k (t)$
\begin{align}
q_k (t)=q_k (0)\cos(\omega_k t)+p_k(0)\sin(\omega_k t)+\alpha_k\int_0^t dt' \sin(\omega_k (t-t'))Q(t'),
\end{align}
and subsequently for $\dot{P}$ (after integration by parts):
\begin{align}
\dot{P}=-\nu Q(t) +\sum_k \left(\frac{\alpha_k^2}{\omega_k} Q(t)-\frac{\alpha_k^2}{\omega_k}\cos(\omega_k t)Q(0)-\frac{\alpha_k^2\nu}{\omega_k}\int_0^t dt' \cos(\omega_k(t-t')) P(t')\right)+\xi (t),
\end{align}
where we introduced the fluctuating force $\xi(t)=\sum_k\left(\alpha_k q_k(0)\cos(\omega_k t)+\alpha_k p_k (0)\sin(\omega_k t)\right)$. One can easily check that the second term in the sum above sums to zero in the continuum limit while the first term is a renormalization of the vibrational frequency with $\tilde{\nu}=\nu-\nu_s$. Using the expressionfor $\alpha_k$ derived in the previous section we can calculate
\begin{align}
\nu_s=\frac{4\Delta k^2}{\omega_{\text{max}}^2\mu\nu m_0}\frac{1}{N+1}\sum_k\cos^2\left(\frac{\pi k}{2}\right)\cos^2\left(\frac{\pi k}{4(N+1)}\right)=\frac{2\Delta k^2}{\omega_{\text{max}}^2\mu\nu m_0}=\frac{\Delta k ^2}{2k_0\mu\nu}=\nu\frac{\Delta k^2}{2k_0k_{\text{M}}}
\end{align}
(for $N\gg 1$).  All together the vibrational relaxation can be written as
\begin{align}
\dot{P}=-\tilde{\nu}Q+\xi (t) -(\Gamma \ast P)(t),
\end{align}
where $\ast$ denotes the convolution $(\Gamma\ast P)(t)=\int_0^\infty dt' \Gamma(t-t' )P(t')$ and
\begin{align}
\Gamma(t)=\sum_k \frac{\alpha_k^2\nu}{\omega_k}\cos(\omega_k t)\Theta(t)=2\nu\nu_s\left[\frac{1}{N+1}\sum_k\cos^2\left(\frac{\pi k}{2}\right)\cos^2\left(\frac{\pi k}{4(N+1)}\right)\cos(\omega_k t)\Theta(t)\right].
\end{align}
In the continuum limit this can be approximated by
\begin{align}
\Gamma(t)=\frac{2\nu\nu_s}{\omegamax}\frac{J_1(\omegamax|t|)}{|t|}\Theta(t),
\end{align}
where $J_n(x)$ are the Bessel functions of first kind. The random force has zero average but non-vanishing correlations
\begin{align}
\braket{\xi(t)\xi(t')}=\sum_k \frac{\alpha_k^2}{2}\left(\coth\left(\frac{\beta\omega_k}{2}\right)\cos(\omega_k(t-t'))-i\sin(\omega_k (t-t'))\right),
\end{align}
where we denote by $\braket{\bullet}=\mathrm{Tr}\left[\bullet\,\rho_{\mathrm{th}}\right]$ the thermal average. The commutator reads
\begin{align}
[\xi(t),\xi(t')]=-i\sum_k\alpha_k^2\sin(\omega_k (t-t')).
\end{align}
In the continuum limit $N\to\infty$ (using the prescription $\sum_k\to\frac{L}{\pi}\int dq\to \int d\omega n(\omega) $ where $\frac{\pi k }{2N+1}=\frac{q L}{2N+1}=qa$) we can rewrite the correlation function as
\begin{align}
\braket{\xi (t)\xi (t')}&=\int_0^{\omega_{\text{max}}}d\omega n(\omega)\frac{\alpha(\omega)^2}{2}\left[\coth\left(\frac{\beta\omega}{2}\right)\cos\left(\omega (t-t')\right)-i\sin\left(\omega(t-t')\right)\right]=\\\nonumber
&=\int_{-\omega_{\text{max}}}^{\omega_{\max}}d\omega n(\omega)\frac{\alpha(\omega)^2}{4}\left[\coth\left(\frac{\beta\omega}{2}\right)+1\right] e^{-i\omega(t-t')}.
\end{align}
The density of states in 1D is given by
\begin{align}
n(\omega)=\frac{L}{\pi}\frac{1}{d\omega/dq}=\frac{2L}{\pi a \sqrt{\omega_{\text{max}}^2-\omega^2}}.
\end{align}
Multiplying this with the vibron-phonon coupling in frequency domain gives
\begin{align}
n(\omega)\alpha(\omega)^2=\frac{4\omega\nu_s}{\pi \omega_{\text{max}}^2}\sqrt{\omega_{\text{max}}^2-\omega^2}.
\end{align}
Using this we can write the correlation function as
\begin{align}
\braket{\xi (t)\xi (t')}&=\frac{1}{2\pi}\int_{-\infty}^{\infty} d\omega \frac{\sqrt{\omega_{\text{max}}^2-\omega^2}}{\omega_{\text{max}}}\Theta(\omega_{\text{max}}-\omega)\Theta(\omega_{\text{max}}+\omega)\frac{\Gamma_{\text{m}}\omega}{\nu}\left(\coth\left(\frac{\beta\omega}{2}\right)+1\right)e^{-i\omega (t-t')},
\end{align}
 where we denote by $\Gamma_{\text{m}}$ the Markovian decay rate $\Gamma_{\text{m}}=\frac{2\nu\nu_s}{\omega_{\text{max}}}$. We note that $\Gamma_{\text{m}}$ actually increases with $\omega_{\text{max}}$ (in our derivation we assumed the reduced mass of the molecule to be equal to the lattice atoms $\mu\approx m_0$): $\Gamma_{\text{m}}=\frac{\Delta k^2}{4 k_0^2}\omega_{\text{max}}$. We can see that we obtain an effective frequency-dependent decay rate $\Gamma_r(\omega)=\frac{\Gamma_{\text{m}}\sqrt{\omega_{\text{max}}^2-\omega^2}}{\omega_{\text{max}}}\Theta(\omega_{\text{max}}-\omega)\Theta(\omega_{\text{max}}+\omega)$ (the real part of the Fourier transform of $\Gamma(t)$). In the limit $\omega_{\text{max}}\to \infty$ this becomes the standard result for a harmonic oscillator in a thermal bath
\begin{align}
\braket{\xi (t)\xi (t')}&=\frac{1}{2\pi}\int_{-\infty}^{\infty} d\omega \frac{\Gamma_{\text{m}}\omega}{\nu}\left(\coth\left(\frac{\beta\omega}{2}\right)+1\right)e^{-i\omega (t-t')}.
\end{align}

\section{Collective vibrational relaxation}
\label{collectivevibrationalrelaxation}

We now consider two molecular impurities inside the 1D chain located at positions $N+1+j$ and $N+1-j$ (distance $2j$). Again we assume that the presence of the two molecules does not significantly change the bulk modes. The Hamiltonian reads:
\begin{align}
H=\frac{\nu_1 P_1^2}{2}+\frac{\nu_1 Q_1^2}{2}+\frac{\nu_2 P_2^2}{2}+\frac{\nu_2 Q_2^2}{2}+\sum_k\left( \frac{\omega_k}{2}p_k^2+\frac{\omega_k}{2}q_k^2- \alpha_{k,1}q_k Q_1-\alpha_{k,2}q_k Q_2\right).
\end{align}
From this we can calculate equations of motion for the molecular coordinates
\begin{align}
\dot{Q}_{1}&=\nu_1 P_1,\\
\dot{Q}_{2}&=\nu_2 {P}_{2},\\
\dot{P}_1&=-\nu_1Q_1+\sum_k \alpha_{k,1}q_k ,\\
\label{prm1}
\dot{P}_{2}&=-\nu_2Q_2+\sum_k \alpha_{k,2}q_k,
\end{align}
as well as for the bath degrees of freedom
\begin{align}
\dot{q}_k &= \omega_k p_k, \\
\dot{p}_k &=-\omega_k q_k +\alpha_{k,1}Q_1 +\alpha_{k,2}Q_2.
\end{align}
Following the same procedure as in Appendix \ref{browniansection} we can solve for $q_k$ to obtain
\begin{align}
q_k(t)=\cos(\omega_k t)q_k(0)+\sin(\omega_k t)p_k(0)+\int_0^t dt' \sin(\omega_k(t-t'))\alpha_{k,1}Q_1(t')+\int_0^t dt' \sin(\omega_k (t-t'))\alpha_{k,2}Q_2(t').
\end{align}
Plugging this into the equations of motion for the system variables we obtain
\begin{align}
\dot{P}_1&=-\nu_1 Q_1 + \xi_1(t)+\sum_k\left(\int_0^t dt' \alpha_{k,1}^2\sin(\omega_k(t-t'))Q_1(t')+\int_0^t dt' \alpha_{k,1}\alpha_{k,2}\sin(\omega_k(t-t'))Q_2(t')\right),\\
\dot{P}_{2}&=-\nu_2 Q_2 + \xi_2(t)+\sum_k\left(\int_0^t dt' \alpha_{k,1}\alpha_{k,2}\sin(\omega_k(t-t'))Q_1(t')+\int_0^t dt' \alpha_{k,2}^2\sin(\omega_k(t-t'))Q_2(t')\right),
\end{align}
where we introduced the input noises $\xi_{1/2}(t) = \sum_k \alpha_{k,{1/2}}\left(\cos(\omega_k t)q_k (0)+\sin(\omega_k t)p_k (0)\right)$. Again integrating by parts we find
\begin{subequations}
\label{fulleqs}
\begin{align}
\dot{P}_1&=-\nu_1 Q_1 (t) +\xi_1(t)+\sum_k \Bigl( \frac{\alpha_{k,1}^2}{\omega_k}Q_1(t)-\frac{\alpha_{k,1}^2}{\omega_k}\cos(\omega_k t)Q_1 (0)-\int_0^t dt' \frac{\alpha_{k,1}^2}{\omega_k}\cos(\omega_k(t-t'))\nu_1P_1 \\\nonumber
&+\frac{\alpha_{k,1}\alpha_{k,2}}{\omega_k}Q_2 (t)-\frac{\alpha_{k,1}\alpha_{k,2}}{\omega_k}\cos(\omega_k t)Q_2 (0)-\int_0^t dt' \frac{\alpha_{k,1}\alpha_{k,2}}{\omega_k}\cos(\omega_k(t-t'))\nu_2 P_2\Bigr),\\
\dot{P}_{2}&=-\nu_2 Q_2 (t) +\xi_2(t)+\sum_k \Bigl( \frac{\alpha_{k,2}^2}{\omega_k}Q_2(t)-\frac{\alpha_{k,2}^2}{\omega_k}\cos(\omega_k t)Q_2 (0)-\int_0^t dt' \frac{\alpha_{k,2}^2}{\omega_k}\cos(\omega_k(t-t'))\nu_2P_2 \\\nonumber
&+\frac{\alpha_{k,1}\alpha_{k,2}}{\omega_k}Q_1 (t)-\frac{\alpha_{k,1}\alpha_{k,2}}{\omega_k}\cos(\omega_k t)Q_1 (0)-\int_0^t dt' \frac{\alpha_{k,1}\alpha_{k,2}}{\omega_k}\cos(\omega_k(t-t'))\nu_1 P_1\Bigr).
\end{align}
\end{subequations}
The two noise terms are correlated according to
\begin{align}
\braket{\xi_1 (t) \xi_2 (t')}=\sum_k \frac{\alpha_{k,1}\alpha_{k,2}}{2}\left(\coth\left(\frac{\beta\omega_k}{2}\right)\cos(\omega_k (t-t'))-i\sin(\omega_k (t-t'))\right).
\end{align}
We now still have to determine the coupling coefficients $\alpha_{k,1}$ and $\alpha_{k,2}$. Assuming that the two molecules are placed at $q_{N+1-j}$ and $q_{N+1+j}$, the couplings are then determined by the neighboring atoms $q_{N+1-j\pm1}$ and $q_{N+1+j\pm1}$:
\begin{align}
q_{\pm 1}=\sqrt{\frac{1}{N+1}}\sum_k \sin\left(\frac{\pi k (N+1-j\pm1)}{2N+2}\right) q_k , \quad q_{\pm 2}=\sqrt{\frac{1}{N+1}}\sum_k \sin\left(\frac{-\pi k(N+1+j\pm1)}{2N+2}\right) q_k,
\end{align}
where we denote the neighboring couplings for the first molecule by $\pm 1$ and for the second molecule by $\pm 2$. With this we can calculate the coupling coefficients (using that $\sin\alpha-\sin\beta = 2\cos(\frac{\alpha+\beta}{2})\sin(\frac{\alpha-\beta}{2})$):
\begin{align}
\alpha_{k,1}&=2\Delta k_1 x_{\text{zpm},1}u_{\text{zpm}}^{(k)}\sqrt{\frac{1}{N+1}}\cos\left(\frac{\pi k(N+1-j)}{2N+2}\right)\sin\left(\frac{\pi k}{2N+2}\right),\\
\alpha_{k,2}&=2\Delta k _2 x_{\text{zpm},2}u_{\text{zpm}}^{(k)}\sqrt{\frac{1}{N+1}}\cos\left(\frac{\pi k(N+1+j)}{2N+2}\right)\sin\left(\frac{\pi k}{2N+2}\right).
\end{align}
With this we obtain for the product $\alpha_{k,1}\alpha_{k,1}$ (using that $\cos(\alpha)\cos(\beta)=[\cos(\alpha+\beta)+\cos(\alpha-\beta)]/{2}$):
\begin{align}
\alpha_{k,1}\alpha_{k,2}&=2\Delta k_1 \Delta k_2 x_{\text{zpm},1}x_{\text{zpm},2}(u_{\text{zpm}}^{(k)})^2\left(\frac{1}{N+1}\right)\left(\cos\left(\frac{\pi k j}{N+1}\right)+\cos\left({\pi k}\right)\right)\sin^2\left(\frac{\pi k}{2N+2}\right)=\\\nonumber
&= \frac{2}{k_0}\sqrt{\nu_{s,1}\nu_{s,2}}\left(u_{\text{zpm}}^{(k)}\right)^2\left(\frac{1}{N+1}\right)\left(\cos\left(\frac{\pi k j}{N+1}\right)+\cos\left({\pi k}\right)\right)\sin^2\left(\frac{\pi k}{2N+2}\right).
\end{align}
We can finally write Eqs.~(\ref{fulleqs}) as (neglecting the terms that sum to zero in the continuum limit)
\begin{subequations}
\label{collectivenonmark}
\begin{align}
\dot{P}_1=-\tilde{v}_1 Q_1-\Omega Q_2-(\Gamma_1\ast P_1)-(\Gamma_{12}\ast P_2)+\xi_1,\\
\dot{P}_2=-\tilde{v}_2 Q_2 -\Omega Q_1-(\Gamma_2\ast P_2)-(\Gamma_{21}\ast P_1)+\xi_2,
\end{align}
\end{subequations}
where we defined $\Omega=\sum_k \frac{\alpha_{k,1}\alpha_{k,2}}{\omega_k}$, $\Gamma_{12}(t)=\sum_k \frac{\alpha_{k,1}\alpha_{k,2}}{\omega_k}\nu_2\cos(\omega_k t)\Theta(t)$, $\Gamma_{21}(t)=\sum_k \frac{\alpha_{k,1}\alpha_{k,2}}{\omega_k}\nu_1\cos(\omega_k t)\Theta(t)$.
Let us for simplicity assume two identical molecules $\nu_1=\nu_2$. We can write the collective decay as
\begin{align}
\Gamma_{12}(t)=\Gamma_{21}(t)=\frac{2\nu\nu_s}{N+1}\sum_k \cos\left(\frac{\pi k (N+1-j)}{2(N+1)}\right)\cos\left(\frac{\pi k (N+1+j)}{2(N+1)}\right)\cos^2\left(\frac{\pi k }{4(N+1)}\right)\cos(\omega_k t)\Theta(t).
\end{align}
In the continuum limit this can be approximated by
\begin{align}
\Gamma_{12}(t)=\Gamma_{21}(t)=\Gamma_{\text{m}}4j\frac{J_{4j}(\omega_{\text{max}}|t|)}{|t|}\Theta(t).
\end{align}
Taking the Fourier transform of Eqs.~(\ref{collectivenonmark}) we obtain a set of algebraic equations (again assuming identical molecules)
\begin{subequations}
\begin{align}
-i\omega P_1(\omega) &= -\tilde{\nu} Q_1(\omega) - \Omega Q_2(\omega) - \sqrt{2\pi}\Gamma_1(\omega)P_1(\omega) - \sqrt{2\pi}\Gamma_{12}(\omega)P_2(\omega) + \xi_1(\omega), \\
-i\omega P_2(\omega) &= -\tilde{\nu} Q_2(\omega) - \Omega Q_1(\omega) - \sqrt{2\pi}\Gamma_2(\omega)P_2(\omega) - \sqrt{2\pi}\Gamma_{12}(\omega)P_1(\omega) + \xi_2(\omega).
\end{align}
\end{subequations}
The Fourier transform of $\Gamma_{12}$ can be obtained in two steps. The Fourier transform of the product $\tilde{f}_{4j}(t)\Theta(t)$ where $\tilde{f}_{4j}(t) = f_{4j}(\omega_{\text{max}}t) = \omega_{\text{max}}(J_{4j}(\omega_{\text{max}}t)/(\omega_{\text{max}}t))$, is given by
\begin{subequations}
\begin{align}
\mathcal{F}(\tilde{f}_{4j}(t)\Theta(t))(\omega) &= \frac{\mathcal{F}(\tilde{f}_{4j})(\omega)}{2} - \frac{1}{2} \int^{\infty}_{-\infty}d\omega' \frac{\mathcal{F}(\tilde{f}_{4j})(\omega')}{i\pi(\omega'-\omega)} \\\nonumber
&= \frac{\mathcal{F}(f_{4j})\left( \frac{\omega}{\omega_{\text{max}}} \right)}{2\omega_{\text{max}}} + \frac{i}{2\pi \omega_{\text{max}}} \int^{\infty}_{-\infty}d\omega' \frac{\mathcal{F}( f_{4j})\left( \frac{\omega'}{\omega_{\text{max}}}\right)}{(\omega'-\omega)},
\end{align}
\end{subequations}
where we have used that $\mathcal{F}(\tilde{f}_{4j})(\omega) = (1/\omega_{\text{max}})\mathcal{F}(f_{4j})(\omega/\omega_{\text{max}})$.
Since the Fourier transform of the term containing the Bessel function results in
\begin{subequations}
\begin{align}
\mathcal{F}\left( \frac{J_{n}(t)}{t} \right)(\omega) &= \sqrt{\frac{2}{\pi}}\frac{i(-i)^n}{4j}U_{n-1}(\omega)\sqrt{1-\omega^2}\text{rect}\left(\frac{\omega}{2} \right),
\end{align}
\end{subequations}
where $U_{n-1}$ is a Chebyshev polynomial of the second kind, we obtain for the collective decay component
\begin{subequations}
\begin{align}
\mathcal{F}(\Gamma_{12})(\omega) &= \frac{1}{\sqrt{2\pi}}\left[i\Gamma_{\text{m}}U_{4j-1}\left(\frac{\omega}{\omega_{\text{max}}} \right)\frac{\sqrt{\omega_{\text{max}}^2-\omega^2}}{\omega_{\text{max}}}\text{rect}\left(\frac{\omega}{2\omega_{\text{max}}} \right) \right. \\\nonumber
& - \left. \frac{\Gamma_{\text{m}}}{\pi} \int^{1}_{-1}dy U_{4j-1}(y)\frac{\sqrt{1-y^2}}{y-x} \right],
\end{align}
\end{subequations}
where $y = \omega'/\omega_{\text{max}}$ and $x = \omega/\omega_{\text{max}}$. For $-1 < x < 1$ meaning that $-\omega_{\text{max}} < \omega < \omega_{\text{max}}$ we have
\begin{subequations}
\begin{align}
\int^{1}_{-1}dy U_{4j-1}(y)\frac{\sqrt{1-y^2}}{y-x} = -\pi T_{4j}(y),
\end{align}
\end{subequations}
with $T_n$ being a Chebyshev polynomial of the first kind \cite{abramowitz1972handbook}, we obtain
\begin{subequations}
\begin{align}
\mathcal{F}(\Gamma_{12})(\omega) &= \frac{1}{\sqrt{2\pi}}\left[i\Gamma_{\text{m}}U_{4j-1}\left(\frac{\omega}{\omega_{\text{max}}} \right)\frac{\sqrt{\omega_{\text{max}}^2-\omega^2}}{\omega_{\text{max}}}\text{rect}\left(\frac{\omega}{2\omega_{\text{max}}} \right) + \Gamma_{\text{m}}T_{4j}\left(\frac{\omega}{\omega_{\text{max}}} \right) \right],
\end{align}
\end{subequations}
and with it
\begin{subequations}
\begin{align}
\mathcal{F}(\Gamma_{12}\ast P_i)(\omega) &= \left[i\Gamma_{\text{m}}U_{4j-1}\left(\frac{\omega}{\omega_{\text{max}}} \right)\frac{\sqrt{\omega_{\text{max}}^2-\omega^2}}{\omega_{\text{max}}}\text{rect}\left(\frac{\omega}{2\omega_{\text{max}}} \right) + \Gamma_{\text{m}}T_{4j}\left(\frac{\omega}{\omega_{\text{max}}} \right) \right]P_i(\omega).
\end{align}
\end{subequations}
In the case $\omega_{\text{max}} \rightarrow \infty$, we obtain $\mathcal{F}(\Gamma_{12}\ast P_i)(\omega) \approx \Gamma_\text{m} P_i(\omega)$ which results in the equations of motion
\begin{subequations}
\begin{align}
\dot{P}_{1} &\approx -\tilde{\nu} Q_1 - \Omega Q_2 - \Gamma_{\text{m}} P_1 - \Gamma_{\text{m}} P_2 + \xi_1, \\
\dot{P}_{2} &\approx -\tilde{\nu} Q_2 - \Omega Q_1 -\Gamma_{\text{m}} P_2 - \Gamma_{\text{m}} P_1 + \xi_2.
\end{align}
\end{subequations}
Considering the coordinates $Q_+ = Q_1+Q_2$, $P_+ = P_1+P_2$ and $ Q_- = Q_{1} - Q_{2}$, $P_- = P_{1} - P_{2}$, we obtain equations of motion for two independent oscillators
\begin{subequations}
\begin{align}
\dot{Q_-} &= \tilde{\nu} P_- ,\\
\dot{P_-} &\approx -(\tilde{\nu}-\Omega)  Q_- + \xi_1 - \xi_2 ,\\
\dot{Q}_+ &= \tilde{\nu}P_+, \\
\dot{P}_+ &\approx -(\tilde{\nu}+\Omega) Q_+ -2\Gamma_\text{m} P_+ + \xi_1 + \xi_2,
\end{align}
\end{subequations}
where the first one is protected from vibrational decay.

\section{Fundamental vibron-phonon processes}
\label{vibronphonon}
To investigate the evolution of a single vibron state we have to develop an optimal framework first.
Starting with the Hamiltonian
\begin{eqnarray}
\label{Eq1}
H &=& \nu b^{\dagger}b + \sum_{k=1}^{N} \omega_{k} c_{k}^{\dagger}c_{k} -\sum_{k=1}^{N} \alpha_{k}(b+b^{\dagger})(c_{k} + c_{k}^{\dagger}),
\end{eqnarray}
we go to the interaction picture following
\begin{eqnarray}
\label{Eq2}
\nonumber
\tilde{H} &=& UHU^{\dagger} -iU\partial_{t}U^{\dagger} \\
&=& -\sum_{k=1}^{N} \alpha_{k} \left( e^{-i\nu t}b + e^{i\nu t}b^{\dagger}\right) \left(e^{-i\omega_{k} t}c_{k} + e^{i\omega_{k} t}c_{k}^{\dagger} \right),
\end{eqnarray}
where $U = e^{iH_{0}t}$ and $H_{0} = \nu b^{\dagger}b + \sum_{k=1}^{N} \omega_{k} c_{k}^{\dagger}c_{k}$. With the initial state in the Schr\"odinger picture given by $\ket{1_{\nu}, \text{vac}_{\text{ph}}}$ we obtain $\ket{\phi} = U\ket{1_{\nu}, \text{vac}_{\text{ph}}} = e^{i\nu t}\ket{1_{\nu}, \text{vac}_{\text{ph}}}$.
Following the Schr\"odinger equation $i\partial_{t}\ket{\phi} = \tilde{H}\ket{\phi}$ we acquire the Dyson series
\begin{align}
\label{Eq3}
\ket{\phi(t)} &= \ket{\phi(0)}- i\int_{0}^{t}dt_{1} \tilde{H}(t_1)\ket{\phi(0)} -\int_{0}^{t}dt_1 \int_{0}^{t_1}dt_2 \tilde{H}(t_1)\tilde{H}(t_2)\ket{\phi(0)} + \dots \\\nonumber
& = \sum_{j=0}^{\infty} (i)^j \int_{0}^{t}\int_{0}^{t_1} \dots \int_{0}^{t_{j-1}}dt_{1} \dots dt_{j} \tilde{H}(t_1)\dots \tilde{H}(t_j)\ket{\phi(0)} \\\nonumber
& = \mathcal{T} e^{-i\int_{0}^{t}d\tau \tilde{H}(\tau)}\ket{1_{\nu}, \text{vac}_{\text{ph}}}.
\end{align}
Evaluating the first order of the Dyson series we obtain
\begin{align}
\label{Eq4}
-i\int_{0}^{t}dt_1 \tilde{H}(t_1)\ket{1_{\nu}, \text{vac}_{\text{ph}}} &= \sum_{k=1}^{N}\alpha_{k}\left[\left(\frac{e^{i(\omega_{k}- \nu)t-1}}{\omega_{k} - \nu} \right)\ket{0_{\nu},1_k} + \sqrt{2}\left(\frac{e^{i(\omega_{k}+\nu)t-1}}{\omega_{k} + \nu} \right)\ket{2_{\nu},1_k} \right].
\end{align}
In the case where we have a resonant component $\omega_{j} = \nu$ with $j \in \{1,\dots,N\}$ and since $\lim_{\omega \rightarrow 0} (e^{i\omega t}-1)/\omega = it$ we obtain
\begin{align}
\label{Eq5}
-i\int_{0}^{t}dt_1 \tilde{H}(t_1)\ket{1_{\nu}, \text{vac}_{\text{ph}}} &= i\alpha_{j}t\ket{0_{\nu},1_j} + \sum_{k \neq j} \alpha_{k}\left(\frac{e^{i(\omega_{k}- \nu)t-1}}{\omega_{k} - \nu} \right)\ket{0_{\nu},1_k} \\\nonumber
& +\sum_{k=1}^{N}\alpha_{k}\sqrt{2}\left(\frac{e^{i(\omega_{k}+\nu)t-1}}{\omega_{k} + \nu} \right)\ket{2_{\nu},1_k},
\end{align}
which can emerge in the case where $\nu \in \{0,\dots, \omega_{\text{max}} \}$. No resonance condition can be fulfilled in the case where $\nu > \omega_{\text{max}}$ and the off resonant terms of Eq.~(\ref{Eq4}) describe the dynamics.
Besides the single phonon resonances in the case $\nu \leq \omega_{\text{max}}$ also weak multi-phonon resonances can be found. For our initial state with one excitation in the vibrational state these terms can be found starting from the third order term of the Dyson expansion. For example for $\omega_{j_1} + \omega_{j_2} + \omega_{j_3} -\nu = 0$ where $j_1 \neq j_2 \neq j_3$ we obtain
\begin{align}
\label{Eq6}
& -i\int_{0}^{t}\int_{0}^{t_1}\int_{0}^{t_2} dt_{1}dt_{2}dt_{3} \alpha_{j_1}\alpha_{j_2}\alpha_{j_3} e^{i[(\omega_{j_1}-\nu)t_1 + (\omega_{j_2}+\nu)t_2 + (\omega_{j_3}-\nu)t_3]}\ket{0_{\nu},1_{j_1},1_{j_2},1_{j_3}} \\\nonumber
& = -i\alpha_{j_1}\alpha_{j_2}\alpha_{j_3}\left[ \frac{i\left(e^{i(\omega_{j_1} + \omega_{j_2} + \omega_{j_3} -\nu)t}-1\right)}{(\omega_{j_1} + \omega_{j_2} + \omega_{j_3} -\nu)(\omega_{j_2} + \omega_{j_3})(\omega_{j_3}-\nu)} -\frac{i\left(e^{i(\omega_{j_1} -\nu)t}-1\right)}{(\omega_{j_1}-\nu)(\omega_{j_2} + \omega_{j_3})(\omega_{j_3}-\nu)} \right. \\\nonumber
& \left. -\frac{i\left(e^{i(\omega_{j_1} + \omega_{j_2})t}-1\right)}{(\omega_{j_1}+\omega_{j_2})(\omega_{j_2} + \nu)(\omega_{j_3}-\nu)} + \frac{i\left(e^{i(\omega_{j_1}-\nu)t}-1\right)}{(\omega_{j_1}-\nu)(\omega_{j_2} + \nu)(\omega_{j_3}-\nu)} \right]\ket{0_{\nu},1_{j_1},1_{j_2},1_{j_3}} \\\nonumber
& = \frac{i\alpha_{j_1}\alpha_{j_2}\alpha_{j_3}t}{(\omega_{j_2} + \omega_{j_3})(\omega_{j_3}-\nu)} \ket{0_{\nu},1_{j_1},1_{j_2},1_{j_3}}  + \dots .
\end{align}
These terms are small with respect to the resonances starting in the first order in the case where $\nu \leq \omega_{\text{max}}$ and in total are comparable to the off resonant terms.
\section{Electron-phonon coupling}
\label{elphcoupling}

\begin{figure}[H]
\center
\includegraphics[width=0.6\columnwidth]{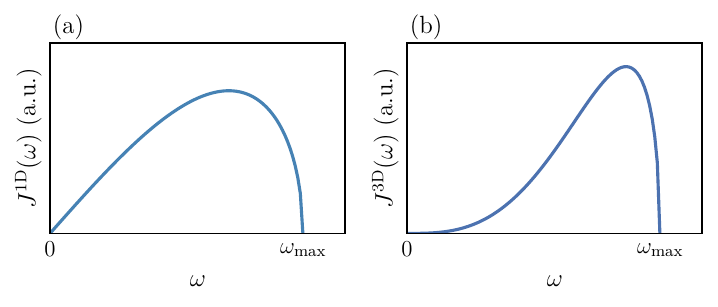}
\caption{\emph{Spectral densities for electron-phonon coupling}. Comparison between spectral densities $J(\omega)$ in a) 1D and b) 3D.}
\label{PlotsSM2}
\end{figure}

The electron-phonon interaction can be diagonalized by means of the polaron transformation $\mathcal{U}=\ket{g}\bra{g}+\mathcal{D}^\dagger\ket{e}\bra{e}$ (additionally leads to a renormalization of the electronic transition frequency $\omega_0-\sum_k\lambda_k^2\omega_k$). This is a collective transformation where $\mathcal{D}^\dagger=\prod_k \mathcal{D}_k^\dagger=e^{-\sum_k \lambda_k (c_k^\dagger-c_k)}$ is the (adjoint) displacement operator for all lattice modes.
Most of the relevant physics is contained in the expectation values of $\braket{\mathcal{D}_k^\dagger}=\braket{\mathcal{D}_k}=e^{-(\lambda_k^2/2) \coth\left({\beta\omega_k/2}\right)}$ as well as in the correlation function
\begin{align}
\braket{\mathcal{D}_k(t)\mathcal{D}_k^\dagger(t')}&=\braket{e^{-i\sqrt{2}\lambda_kp_k(t)}e^{i\sqrt{2}\lambda_k p_k (t')}}=\braket{e^{-i\sqrt{2}\lambda_k\left(p_k (t)-p_k (t')\right)}}e^{\lambda_k^2\braket{[p_k(t),p_k(t')]}}=\\\nonumber
&=e^{-\lambda_k^2\braket{\left(p_k (t)-p_k (t')\right)^2}}e^{\lambda_k^2\braket{[p_k(t),p_k(t')]}}=e^{-2\lambda_k^2\left(\braket{p_k^2}-\braket{p_k(t)p_k(t')}\right)}=\\\nonumber
&=e^{-\lambda_k^2(1+2\bar{n}_k)}e^{\lambda_k^2\left[(1+2\bar{n}_k)\cos\left(\omega_k(t-t')\right)-i\sin\left(\omega_k(t-t')\right)\right]},
\end{align}
where we made use of the fact that Gaussian states are fully characterized by their second-order moments (and odd moments are vanishing) and $p_k(t)=p_k (0)\cos(\omega_k t)-q_k (0)\sin(\omega_k t)$ with $\braket{q_k^2 (0)}=\braket{p_k^2 (0)}=\bar{n}+\frac{1}{2}$ and $\braket{q_k (0)p_k(0)}=-\braket{p_k (0)q_k (0)}=\frac{i}{2}$. We also notice that $\braket{\mathcal{D}_k(t)\mathcal{D}_k^\dagger(t')}=\braket{\mathcal{D}_k^\dagger (t)\mathcal{D}_k(t')}$. In the continuum limit we have (using that $1+2\bar{n}_k=\coth\left(\beta_k\omega_k /2\right)$):
\begin{subequations}
\begin{align}
\braket{\mathcal{D}^\dagger}&=\prod_k\braket{\mathcal{D}_k^\dagger}=e^{-\frac{1}{2}\int_0^{\omega_\text{max}}d\omega n(\omega){\lambda(\omega)^2}\coth\left(\frac{\beta\omega}{2}\right)},\\
\braket{\mathcal{D}(t)\mathcal{D}^\dagger (t')}&=\braket{\mathcal{D}^\dagger (t)\mathcal{D}(t')}=\prod_k\braket{\mathcal{D}_k (t)\mathcal{D}_k^\dagger (t')}=\\\nonumber
&=e^{\int_0^{\omega_\text{max}}d\omega n(\omega)\lambda(\omega)^2\left[\coth\left(\frac{\beta\omega}{2}\right)\left(\cos\left(\omega(t-t')\right)-1\right)-i\sin\left(\omega(t-t')\right)\right]}.
\end{align}
\end{subequations}
We denote by $J(\omega)=n(\omega)\lambda(\omega)^2\omega^2=\sum_k |\lambda_k\omega_k|^2 \delta(\omega-\omega_k)$ the spectral density of the electron-phonon coupling. From Eq.~(\ref{lambdak}) we can derive for the 1D case:
\begin{align}
n^{\text{1D}}(\omega)\lambda(\omega)^2=\frac{8k_\text{tot}^2 \Delta x_{N+2}^2}{\pi m_0 \omega_{\text{max}}^3}\frac{1}{\omega}\frac{\sqrt{\omegamax^2-\omega^2}}{\omegamax},
\end{align}
which leads to a divergence of the integral at small frequencies due to the $1/\omega$-term. (in 1D the spectral density will always be proportional to $\omega$ for low frequencies). Considering a 3D density of states instead:
\begin{align}
n^{\text{3D}}(\omega)=\frac{V}{(2\pi)^3}\frac{4\pi |q(\omega)|^2}{v_{\text{g}}(\omega)},
\end{align}
with group velocity $v_{\text{g}}=\partial\omega/\partial q$ and $q(\omega)$ the inverted dispersion relation in 3D (we make the Debye assumption and assume a linear dependence between wavevector and frequency). With this we obtain for the spectral density
\begin{align}
J^{\text{3D}}(\omega)=\lambda_{\text{e-ph}}^{\text{3D}}\omega^3\frac{\sqrt{\omegamax^2-\omega^2}}{\omegamax},
\end{align}
where we introduced the 3D electron-phonon coupling constant $\lambda_{\text{e-ph}}^{\text{3D}}=\frac{16}{\pi^2}\frac{k_{\text{tot}}^2\Delta x_{\text{N+2}}^2}{m_0\omegamax^5}$ which has units $[s^2]$. Figure \ref{PlotsSM2} shows the schematic behavior of the spectral densities in 1D and 3D.

\section{Calculation of momentum correlation function}
In the case of non-Markovian decay, the time evolution of the momentum operator for the molecular vibration is given by
\begin{align}
\dot{P}(t)=-\tilde{\nu} Q(t) -(\Gamma \ast P)(t) +\xi(t).
\label{eomtime}
\end{align}
 For the calculation of the momentum correlation function $\braket{P(t)P(t')}$ it is convenient to go to Fourier space $\braket{P(t)P(t')}=\frac{1}{2\pi}\int d\omega\int d\omega' e^{-i\omega t}e^{-i\omega' t'}\braket{P(\omega)P(\omega')}$. The equations of motion for $Q$ and $P$  in Fourier space read
\begin{subequations}
\label{eqnsFourier}
\begin{align}
-i\omega Q(\omega)&= \nu P(\omega),\\
-i\omega P(\omega) &=-\tilde{\nu} Q(\omega) -\Gamma (\omega) P(\omega)+\xi(\omega),
\end{align}
\end{subequations}
where $\Gamma(\omega)$ denotes the Fourier transform of the damping kernel $\Gamma (t)$ and is given by
\begin{align}
\Gamma(\omega)=&\Gamma_{\text{m}}\left[ \frac{\sqrt{\omega_{\text{max}}^2-\omega^2}}{\omega_\text{max}}\right]\Theta(\omega_{\text{max}}-\omega)\Theta(\omega_{\text{max}}+\omega)+i\Gamma_{\text{m}}\left[\frac{\sqrt{\omega^2-\omega_{\text{max}}^2}}{\omega_{\text{max}}}\right]\left(\Theta(-\omega-\omega_{\text{max}})-\Theta(\omega-\omega_{\text{max}})\right)+i\frac{\Gamma_{\text{m}}\omega}{\omega_{\text{max}}}\\\nonumber
=:&\Gamma_r(\omega)+i\Gamma_i(\omega).
\end{align}
From Eqs.~(\ref{eqnsFourier}) one can derive $Q(\omega)=\epsilon(\omega)\xi (\omega)$ and $P(\omega)=\chi(\omega)\xi(\omega)$, with the mechanical susceptibility $\chi(\omega)=-i\frac{\omega}{\nu}\epsilon(\omega)=-i\omega[\nu\tilde{\nu}-\omega^2-i\Gamma(\omega)\omega]^{-1}$. The correlations of the input noise in time domain read:
\begin{align}
\braket{\xi (t)\xi(t')}=\frac{1 }{2\pi}\int_{-\omega_{\text{max}}}^{\omega_{\text{max}}}d\omega\frac{\Gamma_r(\omega)\omega}{\nu} \left(\coth\left(\frac{\beta\omega}{2}\right)+1\right) e^{-i\omega(t-t')}=\frac{1}{2\pi}\int_{-\infty}^{\infty}d\omega  S_\text{th}(\omega)  e^{-i\omega(t-t')},
\end{align}
where we defined the thermal spectrum $S_\text{th}(\omega)={\Gamma_r(\omega) \omega} \left(\coth\left(\frac{\beta\omega}{2}\right)+1\right)/\nu$. With this we obtain for the noise correlation in frequency space
\begin{align}
\braket{\xi(\omega)\xi(\omega')}=\frac{1}{(2\pi)^2}\int_{-\infty}^\infty dt \int_{-\infty}^\infty dt' \int_{-\infty}^\infty d\omega'' e^{-i\omega'' (t-t')} S_{\text{th}}(\omega'') e^{i\omega t}e^{i\omega't'}=S_{\text{th}}(\omega)\delta (\omega+\omega'),
\end{align}
from which we see that the noise is always $\delta$-correlated in frequency domain but colored according to the thermal spectrum $ S_{\text{th}}(\omega)$. Using that $\chi(-\omega)=\chi^*(\omega)$ we obtain (assuming $\nu\approx\tilde{\nu}$):
\begin{align}
\label{corrp}
\braket{P(t)P(t')}=\frac{1}{2\pi}\int_{-\infty}^{\infty}d\omega e^{-i\omega(t-t')}|\chi(\omega)|^2S_{\text{th}}(\omega)=\frac{1}{2\pi\nu}\int_{-\infty}^{\infty}d\omega e^{-i\omega (t-t')}\frac{\Gamma_r(\omega) \omega^3\left[\coth\left(\frac{\beta\omega}{2}\right)+1\right]}{(\nu^2-\omega^2+\Gamma_i(\omega)\omega)^2+\Gamma_r(\omega)^2\omega^2},
\end{align}
and similarly for the position correlation
\begin{align}
\braket{Q(t)Q(t')}=\frac{1}{2\pi}\int_{-\infty}^{\infty}d\omega e^{-i\omega (t-t')}\frac{\Gamma_r(\omega)\nu\omega\left[\coth\left(\frac{\beta\omega}{2}\right)+1\right]}{(\nu^2-\omega^2+\Gamma_i(\omega)\omega)^2+\Gamma_r(\omega)^2\omega^2}.
\end{align}
Let us consider these integrals for the simplest case of Markovian decay $\Gamma_r(\omega)=\Gamma_{\text{m}}$, $\Gamma_i(\omega)=0$ and zero temperature. If the susceptibility is sharply peaked $\Gamma_{\text{m}}\ll\nu$, we can solve these integrals by expanding around the pole $\omega=\nu+\delta$, where $\delta$ is a small frequency shift (valid as long as $\Gamma\ll\nu$ and we only have a cooling sideband). We obtain
\begin{align}
\braket{Q(t)Q(t')}=\braket{P(t)P(t')}&=\frac{1}{2\pi} \int_{-\infty}^{\infty} d\delta e^{-i\nu (t-t')}e^{-i\delta (t-t')}\frac{2\Gamma_{\text{m}}\nu^2}{(-2\delta\nu)^2+\Gamma_{\text{m}}^2\nu^2}=\\\nonumber
&=\frac{1}{\sqrt{2\pi}}e^{-i\nu(t-t')}\int_{-\infty}^\infty d\delta \frac{1}{\sqrt{2\pi}}e^{-i\delta (t-t')}\frac{2\Gamma_{\text{m}}}{4\delta^2+\Gamma_{\text{m}}^2}=\\\nonumber
&=\frac{1}{2}e^{-\left(\frac{\Gamma_{\text{m}}}{2}+i\nu\right)|t-t'|},
\end{align}
where we used that $\mathcal{F}(\frac{a}{a^2+\omega^2})=\sqrt{\frac{\pi}{2}}e^{-a|t|}$. In case of a non-zero temperature we now have two sidebands, a cooling sideband at $\nu$ (proportional to $\bar{n}+1$) and a heating sideband at $-\nu$ (proportional to $\bar{n}$). We thus have to expand around both poles $\omega=\nu+\delta$ and $\omega=-\nu+\delta$. This gives the general temperature-dependent momentum correlation function
\begin{align}
\braket{P(t)P(t')}&=\\\nonumber
=&\frac{1}{2\pi}\left[ \int_{-\infty}^{\infty} d\delta e^{-i\nu (t-t')}e^{-i\delta (t-t')}\frac{\Gamma_{\text{m}}\nu^2\left(2+2\bar{n}\right)}{(-2\delta\nu)^2+\Gamma_{\text{m}}^2\nu^2}+\int_{-\infty}^\infty\ d\delta e^{+i\nu (t-t')}e^{-i\delta (t-t')}\frac{\Gamma_{\text{m}}\nu^2\left(2\bar{n}\right)}{(2\delta\nu)^2+\Gamma_{\text{m}}^2\nu^2}\right]=\\\nonumber
&=\left(\frac{\bar{n}+1}{2}\right)e^{-\left(\frac{\Gamma_{\text{m}}}{2}+i\nu\right)(t-t')}+\frac{\bar{n}}{2}e^{-\left(\frac{\Gamma_{\text{m}}}{2}-i\nu\right)(t-t')}=\\\nonumber
&=\left[\left(\bar{n}+\frac{1}{2}\right)\cos(\nu(t-t'))-\frac{i}{2}\sin(\nu(t-t'))\right]e^{-\frac{\Gamma_{\text{m}}}{2}|t-t'|}.
\end{align}

\section{Calculation of absorption spectrum}
\label{calcabsorptionspectrum}

For the calculation of the absorption spectrum we follow the time dynamics of the polaron operator $\tilde{\sigma}=\mathcal{D}^\dagger (t)\mathcal{B}^\dagger (t)\sigma (t)$ (assuming Markovian relaxation for the vibrational mode):
\begin{align}
\label{polaronequation}
\dot{\tilde{\sigma}}=-\left[\gamma-i\left(\omega_\ell-\tilde{\omega}_0\right)\right]\tilde{\sigma}+\eta_\ell \mathcal{D}^\dagger\mathcal{B}^\dagger+\sqrt{2\gamma}\tilde{\sigma}_{\text{in}} -i\lambda\sqrt{2}\Gamma_{\text{m}} \tilde{\sigma}P+i\lambda \sqrt{2}\tilde{\sigma}\xi,
\end{align}
where we defined $\tilde{\omega}_0-\sum_k \lambda_k^2 \omega_k$ and used that $\dot{\tilde{\sigma}}=\dot{\mathcal{D}}^\dagger (t)\mathcal{B}^\dagger (t)\sigma (t)+\mathcal{D}^\dagger (t)\dot{\mathcal{B}}^\dagger (t)\sigma (t)+\mathcal{D}^\dagger (t)\mathcal{B}^\dagger (t)\dot{\sigma} (t)$. When calculating the time derivative of the displacement operators one has to be careful and use Feynman's rule for exponential operators \cite{wilcox1967exponential, suzuki1997quantum}, e.g.~$\dot{\mathcal{B}}^\dagger (t)=\sqrt{2}i\lambda\mathcal{B}^\dagger \int_0^1 ds e^{-is\sqrt{2}\lambda P}\dot{P}(t) e^{is\sqrt{2}\lambda P}$, since $[P(t),\dot{P} (t)]\neq 0$.
As shown in Ref.~\cite{reitz2019langevin} the last two terms in Eq.~(\ref{polaronequation}) cancel out when taking the average over the vibrational degree of freedom and do not contribute. The expectation value of $\braket{\sigma}$ in steady state can then be derived from
\begin{align}
\braket{\sigma}=\eta_\ell\int_{-\infty}^t dt' e^{-\left[\gamma-i\left(\omega_\ell-\tilde{\omega}_0\right)\right](t-t')}\braket{\mathcal{D}(t)\mathcal{D}^\dagger (t')} \braket{\mathcal{B}(t)\mathcal{B}^\dagger (t')},
\end{align}
where the correlation function for the molecular displacement operators is given by (assuming Markovian vibrational relaxation):
\begin{align}
\label{bbdagger}
\braket{\mathcal{B}(t)\mathcal{B}^\dagger(t')}&=\braket{e^{-i\sqrt{2}\lambda P(t)}e^{i\sqrt{2}\lambda P(t')}}=e^{-2\lambda^2\left(\braket{P^2}-\braket{P(t)P(t')}\right)}\\\nonumber
&=e^{-\lambda^2\left(1+2\bar{n}\right)}e^{\lambda^2\left[(2\bar{n}+1)\cos({\nu}(t-t'))-i\sin({\nu}(t-t'))\right]e^{-\frac{\Gamma_{\text{m}}}{2}(t-t')}}.
\end{align}
Let us first consider the purely vibrational part and ignore the electron-phonon part (corresponds to setting all $\lambda_k=0$). We can rewrite
\begin{align}
&e^{\lambda^2\left[(2\bar{n}+1)\cos({\nu}\tau)-i\sin({\nu}\tau)\right]e^{-\frac{\Gamma_{\text{m}}}{2}\tau}}=e^{\lambda^2\left[e^{-i{\nu}\tau}+2\bar{n}\cos({\nu}\tau)\right]e^{-\frac{\Gamma_{\text{m}}}{2}\tau}}=\sum_{n=0}^\infty\frac{\lambda^{2n}}{n!}e^{-n\frac{\Gamma_{\text{m}}}{2}\tau}\left(e^{-i{\nu}\tau}+2\bar{n}\cos({\nu}\tau)\right)^n\\\nonumber
&=\sum_{n=0}^\infty\frac{\lambda^{2n}}{n!}e^{-n\frac{\Gamma_{\text{m}}}{2}\tau}\sum_{k=0}^n\binom{n}{k}e^{-i{\nu}(n-k)\tau}(2\bar{n})^k\cos^k({\nu}\tau)=\sum_{n=0}^\infty \frac{\lambda^{2n}}{n!}e^{-n(\frac{\Gamma_{\text{m}}}{2}+i{\nu} )\tau}\sum_{k=0}^n \sum_{l=0}^k \binom{n}{k} \binom{k}{l}\bar{n}^ke^{2li{\nu}\tau}.
\end{align}
The expectation value of the coherence in steady state becomes
\begin{align}
\braket{\sigma}&=\eta_\ell\int_{-\infty}^t dt' e^{-\left[\gamma-i(\omega_\ell-\omega_0)\right](t-t')}\braket{\mathcal{B}(t)\mathcal{B}^\dagger (t')}\\\nonumber
&=\eta_\ell\int_{-\infty}^t dt' e^{-\left[\gamma-i(\omega_\ell-\omega_0)\right](t-t')} e^{-\lambda^2(1+2\bar{n})}\sum_{n=0}^\infty\frac{\lambda^{2n}}{n!}e^{-n\left(\frac{\Gamma_{\text{m}}}{2}+i{\nu}\right)(t-t')}\sum_{k=0}^n\sum_{l=0}^k\binom{n}{k}\binom{k}{l}\bar{n}^k e^{i2l{\nu}(t-t')}\\\nonumber
&=\eta_\ell\sum_{n=0}^\infty\frac{\lambda^{2n}}{n!}\sum_{l=0}^n\frac{e^{-\lambda^2(1+2\bar{n})}\sum_{k=l}^n\binom{n}{k}\binom{k}{l}\bar{n}^k}{(\gamma+n\frac{\Gamma_{\text{m}}}{2})-i\left[(\omega_\ell-\omega_0)-(n-2l){\nu}\right]}=\eta_\ell\sum_{n=0}^\infty\frac{\lambda^{2n}}{n!}\sum_{l=0}^n\frac{e^{-\lambda^2(1+2\bar{n})}\binom{n}{l}(\bar{n}+1)^{n-l}\bar{n}^l}{(\gamma+n\frac{\Gamma_{\text{m}}}{2})-i\left[(\omega_\ell-\omega_0)-(n-2l){\nu}\right]},
\end{align}
where we made use of the summation rule $\sum_{k=0}^n\sum_{l=0}^k a_{k,l}=\sum_{l=0}^n\sum_{k=l}^n a_{k,l}$ and of the binomial identity $\sum_{k=l}^n \binom{n}{k}\binom{k}{l}x^k=x^l(1+x)^{n-l}\binom{n}{l}$. The steady-state excited state population (extinction spectrum) in thermal equilibrium with the environment then follows from $\mathcal{P}_{\text{e}}=\braket{\sigma^\dagger\sigma}=\frac{\eta_\ell}{2\gamma}\left(\braket{\sigma}+\braket{\sigma}^*\right)$:
\begin{align}
\mathcal{P}_{\text{e}}=\frac{\eta_\ell^2}{\gamma}\sum_{n=0}^\infty\frac{\lambda^{2n}}{n!}\sum_{l=0}^n\binom{n}{l}\frac{e^{-\lambda^2(1+2\bar{n})}(\bar{n}+1)^{n-l}\bar{n}^l \left(\gamma+n\frac{\Gamma_{\text{m}}}{2}\right)}{(\gamma+n\frac{\Gamma_{\text{m}}}{2})^2+\left[(\omega_\ell-\omega_0)-(n-2l){\nu}\right]^2}.
\end{align}
Using that ${(\bar{n}+1)}/{\bar{n}}=e^{\beta\nu}$ we can also rewrite the displacement correlation function Eq.~\eqref{bbdagger} as
\begin{align}
\braket{\mathcal{B}(t)\mathcal{B}^\dagger (t')}=e^{-\lambda^2(1+2\bar{n})}e^{\lambda^2 e^{-\Gamma_{\text{m}}(t-t')}\sqrt{\bar{n}(\bar{n}+1)}\left[e^{-i\nu(t-t')}e^{\beta\nu/2}+e^{i\nu(t-t')}e^{-\beta\nu/2}\right]}.
\end{align}
The generating function of the \textit{modified} Bessel functions $I_n (x)$ is given by $e^{\frac{1}{2}x(a+1/a)}=\sum_{n=-\infty}^{\infty}I_n (x) a^n$. This yields
\begin{align}
\braket{\mathcal{B}(t)\mathcal{B}^\dagger (t')}=e^{-\lambda^2(1+2\bar{n})} \sum_{n=-\infty}^\infty I_n\left(2\lambda^2\sqrt{\bar{n}(\bar{n}+1)}e^{-\Gamma_{\text{m}}(t-t')}\right)e^{-ni\nu(t-t')}\left(\frac{\bar{n}+1}{\bar{n}}\right)^{n/2}.
\end{align}
For $|x|\ll 1$ one can approximate $I_n(ax)=I_n (x) a^{|n|}$. With this the integral over time can be performed, resulting in the following expression for the absorption spectrum
\begin{align}
{\mathcal{P}_{\text{e}}}=\eta_\ell^2\sum_{n=-\infty}^\infty \frac{f_{\text{FC}}\left(\frac{\bar{n}+1}{\bar{n}}\right)^{n/2}I_n\left(2\lambda^2\sqrt{\bar{n}(\bar{n}+1)}\right)\left(\gamma+|n|\frac{\Gamma_{\text{m}}}{2}\right)/\gamma}{(\gamma+|n|\frac{\Gamma_{\text{m}}}{2})^2+(\omega_\ell-\omega_0-n\nu')^2},
\end{align}
which is similar to the expression known from Huang-Rhys theory \cite{huang1950theory}. Including the phonon modes, the collective phonon mode displacement correlation for $N$ phonon modes can be expressed as (neglecting phonon decay):
\begin{align}
\braket{\mathcal{D}(t)\mathcal{D}^\dagger (t')}=\sum_{n_1,\hdots,n_N}^{\infty}\sum_{l_1,\hdots,l_N}^{n_1,\hdots,n_N}e^{-i\sum_{k=1}^N \left[(n_k-2l_k)\omega_k\right](t-t')}\prod_{k=1}^N \left[ \frac{\lambda^{n_k}}{n_k!}e^{-\lambda_k^2(1+2\bar{n}_k)}\binom{n_k}{l_k}\left(\bar{n}_k+1\right)^{n_k-l_k}\bar{n}_k^{l_k}\right].
\end{align}
Together with the the vibrational modes, the total absorption spectrum can then be expressed in terms of discrete lines as
\begin{align}
\mathcal{P}_{\text{e}}=\frac{\eta_\ell^2}{\gamma}\sum_{n=0}^\infty\sum_{l=0}^n \frac{\lambda^{2n}}{n!}e^{-\lambda^2 (1+2\bar{n})}\binom{n}{l}(\bar{n}+1)^{n-l}\bar{n}^l \sum_{n_1,\hdots,n_N}^{\infty}\sum_{l_1,\hdots,l_N}^{n_1,\hdots,n_N}\frac{\prod_{k=1}^N L_k(n_k) B_k (n_k,l_k)\left(\gamma+n\frac{\Gamma_{\text{m}}}{2}\right)}{\left(\gamma+n\frac{\Gamma_{\text{m}}}{2}\right)^2+\left[(\omega_\ell-\tilde{\omega}_0)-(n-2l)\nu-\sum_k(n_k-2l_k)\omega_k\right]^2},
\end{align}
where we introduced $L_k(n_k)=\frac{\lambda^{2n_k}}{n_k!} e^{-\lambda_k^2(1+2\bar{n}_k)}$ and $B_k(n_k,l_k)=\binom{n_k}{l_k}(\bar{n}_k+1)^{n_k-l_k}\bar{n}_k^{l_k}$.

\section{Off-diagonal electron-vibron coupling}
\label{offdiagonal}
\textcolor{black}{We want to discuss briefly how on can include off-diagonal electron-vibration coupling (e.g.~proportional to $\sigma_x$) with coupling constant $\lambda_{x}$ into the equations of motion for the polaron operator. Such terms can become relevant in the case of nonadiabatic molecular dynamics and can in the simplest form be described by:
\begin{align}
H_{\sigma_x}=\lambda_{x}(b^\dagger+b)\sigma_x=\lambda_{x}(b^\dagger+b)(\sigma^\dagger+\sigma).
\end{align}
The Langevin equation of motion for the vibronic polaron operator $\tilde{\sigma}=\sigma\mathcal{B}^\dagger$ is then given by
\begin{align}
\label{sigmax}
\dot{\tilde{\sigma}}=-(\gamma-i\Delta_\ell)\tilde{\sigma}+\eta_\ell \mathcal{B}^\dagger+\sqrt{2\gamma}\tilde{\sigma}_{\text{in}}-i\lambda_{x}(b^\dagger+b)\mathcal{B}^\dagger(1-2\sigma^\dagger\sigma)-2i\lambda\lambda_{x}\mathcal{B}^\dagger\sigma\sigma^\dagger.
\end{align}
Under the assumption of weak driving ($\braket{\sigma\sigma^\dagger}\approx 1$), this equation can be formally integrated as:
\begin{align}
\braket{\sigma }=(\eta_\ell-2i\lambda\lambda_x)\int_{-\infty}^t dt' e^{-(\gamma-i\Delta_\ell)(t-t')}\braket{\mathcal{B}(t)\mathcal{B}^\dagger(t')}-i\lambda_x\int_{-\infty}^t dt' e^{-(\gamma-i\Delta_\ell)(t-t')}\braket{\mathcal{B}(t)\left(b^\dagger (t')+b(t')\right)\mathcal{B}^\dagger(t')},
\end{align}
which requires the correlation function $\braket{\mathcal{B}(t)\left(b^\dagger (t')+b(t')\right)\mathcal{B}^\dagger(t')}$. From Eq.~(\ref{sigmax}) on can however already estimate the amplitude of the ZPL line transition (other amplitudes can be estimated in a similar way) by expanding the displacement operator as $\mathcal{B}^\dagger(t)=e^{-\lambda^2/2}\sum_{k,l} \frac{(-\lambda)^k}{k!}\frac{\lambda^l}{l!} b^\dagger(t)^k b(t)^{l}$ and averaging over the vibrational ground state $\ket{0_v}$. This gives rise to an effective additional driving of the ZPL line transition with $\eta_\ell-i\lambda\lambda_x$.}

\section{Cavity spectroscopy}

We now consider additional coupling to a cavity mode described by the Jaynes-Cummings Hamiltonian $H_{\text{JC}}=g(a^\dagger \sigma+a \sigma^\dagger)$. We now consider a driving of the cavity mode instead of the molecule. The equations of motion for $a$ and the polaron operator $\tilde{\sigma}$ read
\begin{align}
\dot{a}&=-\left[\kappa-i(\omega_\ell-\omega_c)\right]a-ig\sigma+\sqrt{2\kappa}A_{\text{in}},\\
\dot{\tilde{\sigma}}&=-\left[\gamma-i\left(\omega_\ell-\tilde{{\omega}}_0\right)\right]\tilde{\sigma}-ig \mathcal{D}^\dagger \mathcal{B}^\dagger a+\sqrt{2\gamma}\tilde{\sigma}_{\text{in}},
\end{align}
where we defined $A_{\text{in}}={\eta_c}/\sqrt{{2\kappa}}+a_{\text{in}}$. We can formally integrate the equation for the polaron operator:
\begin{align}
\sigma(t)= -ig \int_{-\infty}^t dt' e^{-\left[\gamma-i\left(\omega_\ell-\tilde{{\omega}}_0\right)\right](t-t')}\left[  \mathcal{D}(t) \mathcal{D}^\dagger(t') \mathcal{B}(t) \mathcal{B}^\dagger(t') a(t')-\frac{\sqrt{2\gamma}}{ig}\mathcal{D}(t)\mathcal{B}(t)\tilde{\sigma}_{\text{in}}(t')\right].
\end{align}
We can now plug this into the equation for $a$ (additionally we average over the vibrational as well as electronic degrees of freedom):

\begin{align}
\dot{a}&=-\left[\kappa-i(\omega_\ell-\omega_c)\right]a-g^2\int_{-\infty}^t dt' e^{-\left[\gamma-i\left(\omega_\ell-\tilde{{\omega}}_0\right)\right](t-t')} \left[\braket{\mathcal{D}(t) \mathcal{D}^\dagger(t')}\braket{ \mathcal{B}(t) \mathcal{B}^\dagger(t')} a(t')\right]+\sqrt{2\kappa}A_{\text{in}}.
\end{align}
Taking the average over the cavity field we obtain
\begin{align}
\braket{\dot{a}}&=-\left[\kappa-i(\omega_\ell-\omega_c)\right]\braket{a}-g^2\int_{-\infty}^\infty dt' e^{-\left[\gamma-i\left(\omega_\ell-\tilde{{\omega}}_0\right)\right](t-t')} \Theta(t-t') \braket{\mathcal{D}(t)\mathcal{D}^\dagger (t')}\braket{ \mathcal{B}(t) \mathcal{B}^\dagger(t')} \braket{a(t')}+\eta_c,
\end{align}
where we assumed factorizability between vibrational, phononic and optical degrees of freedom. We notice that the second term represents a convolution since $\braket{\mathcal{D}(t)\mathcal{D}^\dagger (t')}\braket{ \mathcal{B}(t) \mathcal{B}^\dagger(t')} $ just depends on the time difference $t-t'$. Making the notation $\mathcal H(t-t')=e^{-\left[\gamma-i\left(\omega_\ell-\tilde{{\omega}}_0\right)\right](t-t')} \Theta(t-t')\braket{\mathcal{D}(t)\mathcal{D}^\dagger (t')} \braket{ \mathcal{B}(t) \mathcal{B}^\dagger(t')}$, we obtain in Fourier space
\begin{align}
-i\omega\braket{a(\omega)}&=-\left[\kappa-i(\omega_\ell-\omega_c)\right]\braket{a(\omega)}-g^2 \mathcal H(\omega) \braket{a(\omega)}+\eta_c\delta(\omega).
\end{align}
The electric field amplitude in Fourier space can therefore be expressed as
\begin{align}
\braket{a(\omega)}=\frac{\eta_c\delta(\omega)}{g^2 \mathcal H(\omega)-i\omega+\left[\kappa-i(\omega_\ell-\omega_c)\right]},
\end{align}
and is related to the normalized cavity transmission $\mathcal{T}(\omega)$ via $\mathcal{T}(\omega)=\frac{\braket{A_{\text{out}}(\omega)}}{\braket{A_{\text{in}}(\omega)}}$. This gives
\begin{align}
\mathcal{T}(\omega)=\frac{\kappa}{g^2 \mathcal H(\omega)-i\omega+\left[\kappa-i(\omega_\ell-\omega_c)\right]}.
\end{align}
 Let us focus on the purely vibrational part, i.e., $\mathcal H(t-t')=e^{-\left[\gamma-i\left(\omega_\ell-{\omega}_0\right)\right](t-t')} \Theta(t-t') \braket{ \mathcal{B}(t) \mathcal{B}^\dagger(t')}$. We then obtain for a molecule in thermal equilibrium (assuming Markovian decay)
\begin{align}
\mathcal H(\omega)=\sum_{n=0}^\infty \frac{\lambda^{2n}}{n!}\sum_{l=0}^n \binom{n}{l}\frac{e^{-\lambda^2(1+2\bar{n})}\left(\bar{n}+1\right)^{n-l} \bar{n}^l}{\left(\gamma+n\frac{\Gamma}{2}\right)-i\left[(\omega+\omega_\ell-\omega_0)-(n-2l)\nu\right]},
\end{align}
which simplifies to
\begin{align}
\mathcal H(\omega)=\sum_{n=0}^\infty \frac{\lambda^{2n}}{n!}\frac{e^{-\lambda^2}}{\left(\gamma+n\frac{\Gamma}{2}\right)-i\left[(\omega+\omega_\ell-\omega_0)-n\nu\right]},
\end{align}
in the case of zero temperature $\bar{n}=0$.

\section{Polariton cross-talk}
\label{polcrosstalk}
Let us start with the resonant Holstein-Cummings Hamiltonian for a single vibrational or phononic mode
\begin{align}
H=\left(\omega_0+\lambda^2\nu\right)\sigma^\dagger\sigma+\omega_0 a^\dagger a+\nu b^\dagger b -\lambda\nu(b^\dagger+b)\sigma^\dagger\sigma+g(a^\dagger\sigma+a\sigma^\dagger),
\end{align}
where the cavity is resonant to the bare electronic transition energy $\omega_0$. We diagonalize the Jaynes-Cummings part in the single-excitation subspace by introducing annihilation operators for upper and lower polariton.
\begin{subequations}
\begin{align}
U=\frac{1}{\sqrt{2}}(a+\sigma),\\
L=\frac{1}{\sqrt{2}}(a-\sigma).
\end{align}
\end{subequations}
Inserting this into the Hamiltonian gives
\begin{align}
H=(\omega_0+g+\frac{\lambda^2\nu}{2})U^\dagger U+(\omega_0-g+\frac{\lambda^2\nu}{2})L^\dagger L+\nu b^\dagger b -\frac{\lambda\nu}{2}(b^\dagger+b)\left(U^\dagger U + L^\dagger L\right) +\frac{\lambda\nu}{2}(U^\dagger L+L^\dagger U)(b^\dagger+b).
\end{align}
We can interpret this as two vibrationally-dressed polaritons where the last term describes a coupling between upper and lower polariton. We can perform a polaron transform for both upper $\mathcal{U}_\text{U}=\ket{g}\bra{g}+\mathcal{B}_{\text{pol}}^\dagger \ket{U}\bra{U}$ and lower polaritonic state $\mathcal{U}_\text{L}=\ket{g}\bra{g}+\mathcal{B}_{\text{pol}}^\dagger \ket{L}\bra{L}$ where we introduce the displacement operator $\mathcal{B}_{\text{pol}}^\dagger=e^{-\frac{\lambda}{2}(b^\dagger-b)}$. Using that $\mathcal{U}_\text{U}b\mathcal{U}_\text{U}^\dagger=b+\frac{\lambda}{2}U^\dagger U$ (similarly for $\mathcal{U}_\text{L}$) this gives the transformed Hamiltonian
\begin{align}
H_{\text{pol}}=\omega_+ U^\dagger U + \omega_- L^\dagger L+\nu b^\dagger b +\frac{\lambda\nu}{2}(U^\dagger L + L^\dagger U)(b^\dagger + b)+\frac{\lambda^2\nu}{2}(U^\dagger L +L^\dagger U),
\end{align}
with $\omega_\pm=\omega_0\pm g+\frac{\lambda^2\nu}{4}$. We notice the additional last term which describes a direct polariton-polariton coupling, which is however negligible in the case of large Rabi splitting $2g\gg \lambda^2\nu$. We start with the equations of motion for the individual polaritons
\begin{subequations}
\begin{align}
\frac{dU}{dt}&=-(\gamma_++i\omega_+)U-i\frac{\lambda\nu}{2}L(b^\dagger+b)+\sqrt{\kappa}a_{\text{in}}+\sqrt{\gamma}\sigma_{\text{in}},\\
\frac{dL}{dt}&=-(\gamma_-+i\omega_-)L-i\frac{\lambda\nu}{2}U(b^\dagger+b)+\sqrt{\kappa}a_{\text{in}}-\sqrt{\gamma}\sigma_{\text{in}},
\end{align}
\end{subequations}
as well as the populations
\begin{subequations}
\begin{align}
\frac{d\mathcal{P}_U}{dt}&=-2\gamma_+ \mathcal{P}_U+\lambda\nu\Im\braket{U^\dagger L(b^\dagger+b)},\\
\frac{d\mathcal{P}_L}{dt}&=-2\gamma_- \mathcal{P}_L+\lambda\nu\Im\braket{ L^\dagger U  (b^\dagger+b)},
\end{align}
\end{subequations}
with the hybridized decay rates $\gamma_\pm=\frac{\kappa+\gamma}{2}$. We will calculate the energy transfer rate from upper to lower polariton $\kappa_+$ in a perturbative approach (the transfer from lower to upper polariton is then calculated analogously). To this end, one can assume initial excitation of the upper polariton $U^\dagger U (0)=\mathcal{P}_U (0)$ and consider the transfer to the lower polariton. If the decay rate of the upper polariton $\gamma_+$ is much larger than the transfer rate $\kappa_+$ one can assume that the population of the upper polariton is not modified due to the transfer $\mathcal{P}_U\approx -2\gamma_+ \mathcal{P}_U$ but the population of the lower polariton sees an increase $\dot{\mathcal{P}}_L=-2\gamma_- \mathcal{P}_L+\kappa_{+}\mathcal{P}_U$. We have to calculate
\begin{align}
\braket{L^\dagger U (b^\dagger(t) + b(t))}&=i\frac{\lambda\nu}{2}\int_{-\infty}^t dt' \mathcal{P}_U(0) \braket{\left(b(t')+b^\dagger (t')\right)\left(b(t)+b^\dagger (t)\right)}e^{-(\gamma_--i\omega_-)(t-t')}e^{-(\gamma_+-i\omega_+)t'}e^{-(\gamma_++i\omega_+)t}\\\nonumber
&=i\frac{\lambda\nu}{2}\left[\frac{\bar{n}+1}{\Gamma_{\text{m}}/2+i\left(\omega_+-\omega_--\nu\right)}+\frac{\bar{n}}{\Gamma_{\text{m}}/2+i\left(\omega_+-\omega_-+\nu\right)}\right]\mathcal{P}_U (t),
\end{align}
where we assumed Markovian decay for the vibration and assumed fast vibrational relaxation $\Gamma_{\text{m}}\gg\gamma_{\pm}$. This leads to an energy transfer rate of
\begin{align}
\kappa_{+}=\frac{\lambda^2\nu^2}{4}\Gamma_{\text{m}}\left[\frac{\bar{n}+1}{(\Gamma_{\text{m}}/2)^2+\left(\omega_+-\omega_--\nu\right)^2}+\frac{\bar{n}}{(\Gamma_{\text{m}}/2)^2+\left(\omega_+-\omega_-+\nu\right)^2}\right].
\end{align}
The transfer from lower to upper polariton can then be calculated by assuming initial occupation of the lower polariton from $\braket{U^\dagger L (b^\dagger+b)}$ as
\begin{align}
\kappa_{-}=\frac{\lambda^2\nu^2}{4}\Gamma_{\text{m}}\left[\frac{\bar{n}+1}{(\Gamma_{\text{m}}/2)^2+\left(\omega_+-\omega_-+\nu\right)^2}+\frac{\bar{n}}{(\Gamma_{\text{m}}/2)^2+\left(\omega_+-\omega_--\nu\right)^2}\right].
\end{align}

\end{document}